\documentclass[journal,twoside,web]{ieeecolor}
\usepackage{tmi}
\usepackage{cite}
\usepackage{amsmath,amssymb,amsfonts}
\usepackage{algorithmic}
\usepackage{graphicx}
\usepackage{textcomp}
\usepackage{float}
\usepackage{mhchem}
\usepackage{booktabs}
\usepackage{multirow}
\usepackage{hyperref}
\hypersetup{
colorlinks=true,
linkcolor=red,
citecolor=green
}

\def\BibTeX{{\rm B\kern-.05em{\sc i\kern-.025em b}\kern-.08em
    T\kern-.1667em\lower.7ex\hbox{E}\kern-.125emX}}
\begin{document}

\title{Label-Free Segmentation of COVID-19 Lesions in Lung CT}

\author{Qingsong Yao,~\IEEEmembership{Student Member,~IEEE}, Li Xiao,~\IEEEmembership{Member,~IEEE}, Peihang Liu 

and S. Kevin Zhou,~\IEEEmembership{Fellow,~IEEE}
\thanks{Yao, Xiao and Zhou are with Institute of Computing Technology, Chinese Academy of Sciences. Zhou is corresponding author. Emails: yaoqingsong19@mails.ucas.edu.cn; {xiaoli, zhoushaohua}@ict.ac.cn.} 
\thanks{Liu is with Beijing University of Posts and Telecommunications. Email: phliu@bupt.edu.cn.}

}

\maketitle

\begin{abstract}
Scarcity of annotated images hampers the building of automated solution for reliable COVID-19 diagnosis and evaluation from CT. To alleviate the burden of data annotation, we herein present a label-free approach for segmenting COVID-19 lesions in CT via voxel-level anomaly modeling that mines out the relevant knowledge from normal CT lung scans. Our modeling is inspired by the observation that the parts of tracheae and vessels, which lay in the high-intensity range where lesions belong to, exhibit strong patterns. To facilitate the learning of such patterns at a voxel level, we synthesize `lesions' using a set of simple operations and insert the synthesized `lesions' into normal CT lung scans to form training pairs, from which we learn a normalcy-recognizing network (NormNet) that recognizes normal tissues and separate them from possible COVID-19 lesions. Our experiments on three different public datasets validate the effectiveness of NormNet, which conspicuously outperforms a variety of unsupervised anomaly detection (UAD) methods.
\end{abstract}


\section{Introduction}
\label{sec:introduction}
\IEEEPARstart{T}{he} world has been facing a global pandemic caused by a novel Coronavirus Disease (COVID-19) since December 2019 \cite{covid-situation,covid-review}. According to the report from World Health Organization, COVID-19 has infected over 62 millions people including more than half a million deaths up to November 30 \cite{WHO}. In clinics, real-time reverse-transcription–polymerase-chainreaction (RT-PCR)~\cite{pcr,radioreport1} and the radiological imaging techniques, e.g., X-ray and computed tomography (CT), play a key role in COVID-19 diagnosis and evaluation \cite{covid-review,ct-role}. 

Due to the high spatial resolution and the unique relationship between CT density and lung air content \cite{ct-lung,zhou1,zhou2,zhou3}, CT is widely preferred to recognize and segment the typical signs of COVID-19 infection~\cite{ct-seg1}. Furthermore, segmentation of COVID-19 lesions provides crucial information for quantitative measurement and follow-up assessment \cite{ct-seg2}. As it is time-consuming for experts to go through the 3D CT volumes slice by slice, automatic segmentation is highly desirable in clinical practice \cite{covid-review, ct-seg3}. Recently, deep learning based methods have been proposed for COVID-19 lesion screening \cite{covid-review} and some of them are proved successful for COVID-19 segmentation \cite{ct-seg1,ct-seg2,ct-seg3}.

\begin{table}[h]
 \caption{A summary of public COVID-19 datasets. The quantity is specific to the cases of COVID-19.} 
 \begin{tabular}{llrl} 
  \toprule 
Dataset & Modality & Quantity & Task   \\ 
  \midrule 
COVID-CT\cite{CTdatsetCLS1}             & CT image    & 342            & Diagnosis      \\
SIRM-COVID\cite{sirm}           & 2D CT image    & 340            & Diagnosis      \\
SIRM-Seg\cite{sirm,datasetseg1}  & CT image    & 110            & Segmentation  \\
Radiopedia\cite{datasetseg2,datasetseg1}           & CT volume   & 9              & Segmentation   \\
Coronacase\cite{datasetseg3,junma}            & CT volume   & 20             & Segmentation   \\
Mosmed\cite{mosmed}               & CT volume   & 50             & Diagnosis      \\
BIMCV\cite{BIMCV}        & CT / X-rays & 5381           & Diagnosis      \\
UESTC\cite{tmi-noise}   & CT volume &120 & Segmentation \\
  \bottomrule 
 \end{tabular} 
 \label{public}
\end{table}

Despite such success, they all rely on large-scale well-labeled datasets. However, obtaining such datasets is very difficult due to two related concerns. On the one hand, labeling a 3D CT volume is costly and time-consuming. Often it needs experienced radiologists, who are busy fighting the COVID-19 pandemic and hence lack time for lesion labeling. On the other hand, the COVID-19 lesions not only have a variety of complex appearances such as Ground-Glass Opacity (GGO), reticulation, and consolidation~\cite{radioreport1}, but also have high variations in texture, size, and position. Those diversities raise a great demand for rich annotated datasets. Accordingly, large-scale well-labeled COVID-19 datasets are scarce, which limits the use of Artificial Intelligence (AI) to help fight against COVID-19. As reported in Table \ref{public}, most of the public COVID-19 datasets focus on diagnosis which only have classification information, while only a few of them provide semantic segmentation labels. 
While research attempts~\cite{tmi-inf,tmi-noise,tmi-rapid} have been made to address the challenges, these works, nevertheless, still need annotated images for training purpose. In this paper, we present a \textbf{label-free approach}, requiring no lesion annotation.

\begin{figure*}[t]
\centerline{\includegraphics[width=1.3\columnwidth]{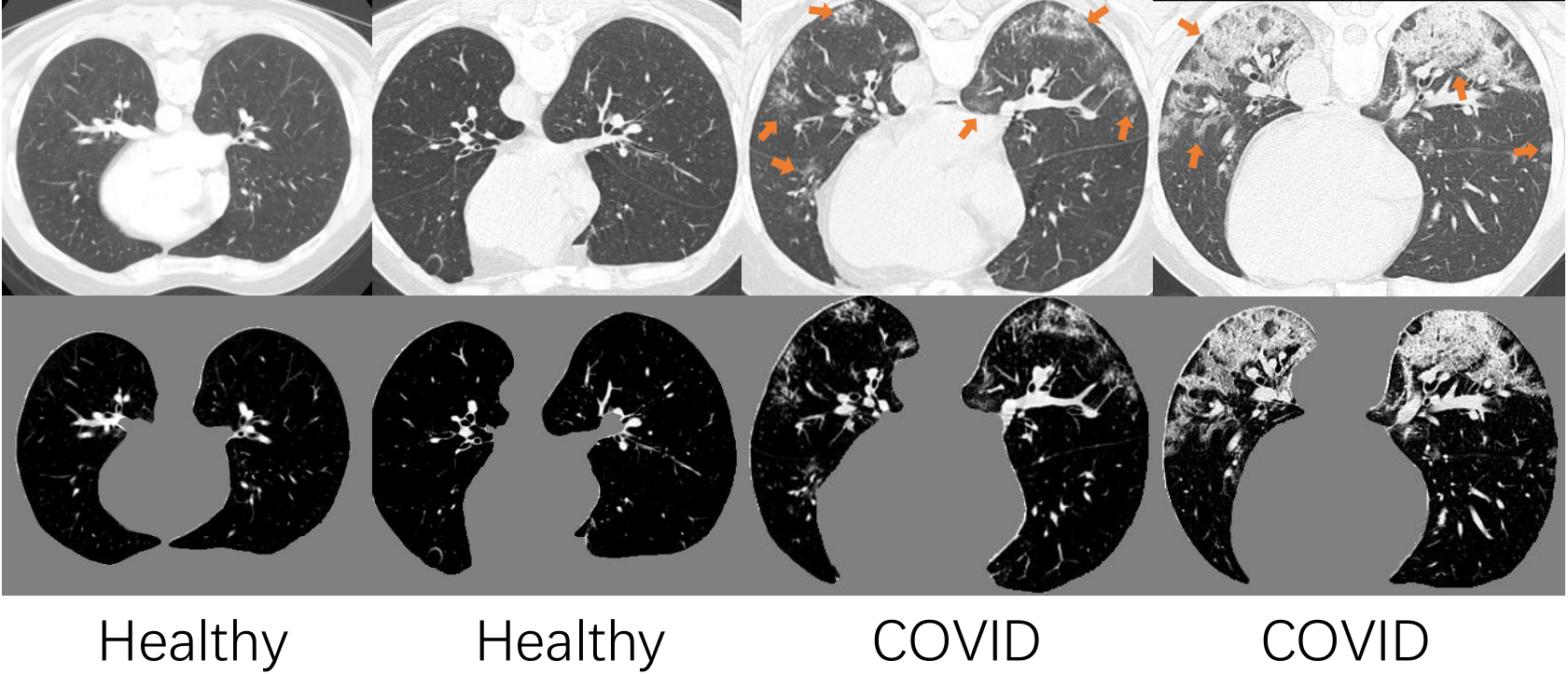}(a)
\includegraphics[width=0.6\columnwidth]{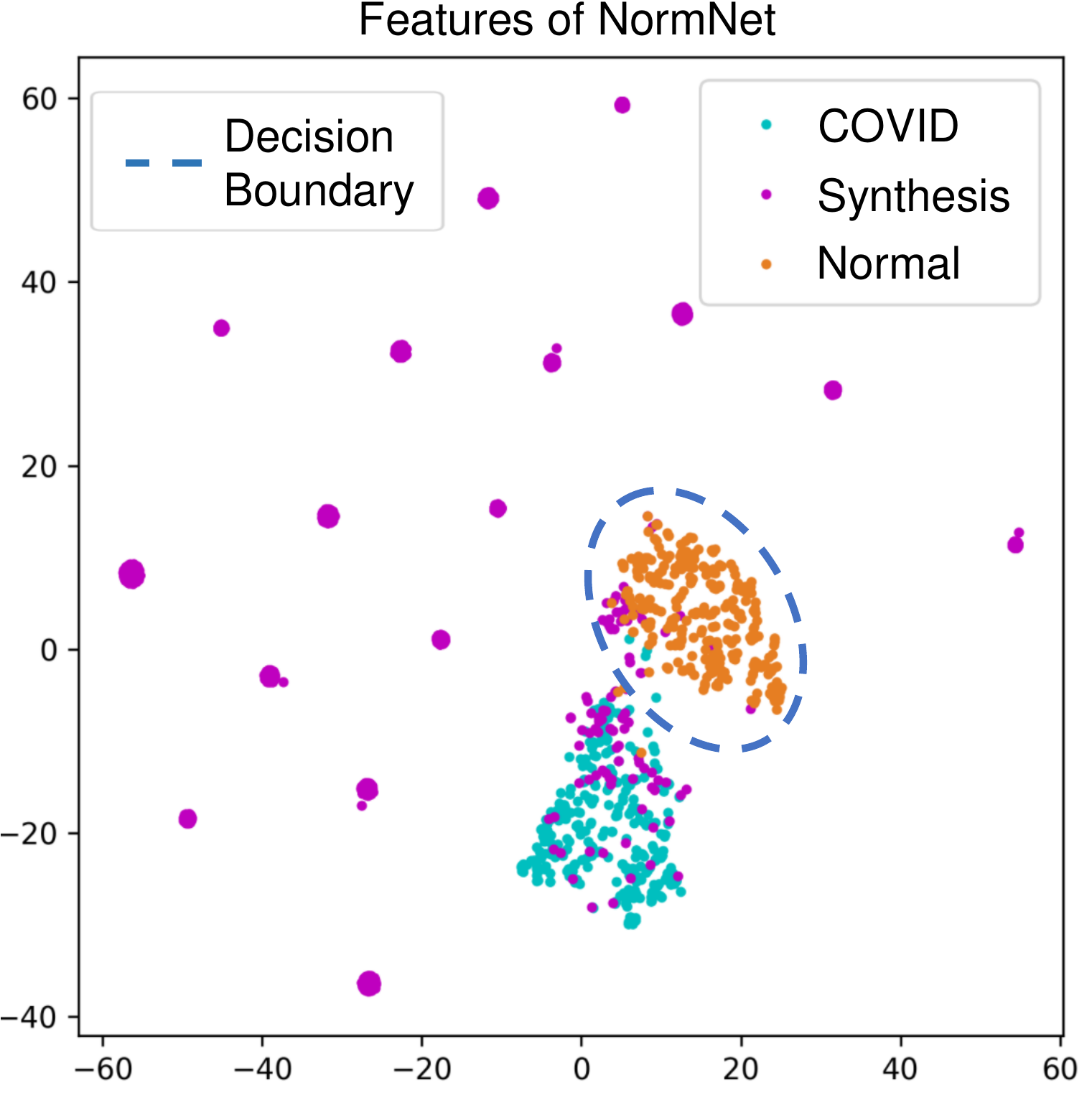}(b)}
\caption{(a) Healthy and COVID-19 lung CT images (top) and its corresponding thorax area (bottom), clipped with an Hounsfield unit (HU) range of $[-800,100]$ and scaled to $[0,1]$. (b) The visualization of 2D t-SNE of the features from the last layer of NormNet for COVID-19 lesions, synthetic `lesions' and normal contexts. We build a rich synthetic `lesion' library, which serves as a \textbf{superset} relative to the COVID-19 lesions. The NormNet learns a tight decision boundary between normal textures and the diverse `lesions', which can further be used to segment COVID-19 lesions.}
\vspace{-3mm}
\label{fig:intro}
\end{figure*}

Although it is very difficult to build a large well-labeled COVID-19 dataset, collecting a large-scale normal CT volume dataset is much easier. It is also interesting to notice that the patterns of normal lungs are regular and easy to be modeled. The thorax of a normal person consists of large areas of air and a few tissues (such as tracheae and vessels \cite{ct-lung}), which can be clearly distinguished by CT intensity \cite{ct-lung}. As shown in Fig. \ref{fig:intro}(a), the air region is usually displayed as black background, with its Hounsfield unit (HU) value around -1000 \cite{ct-lung}. Meanwhile, the tissue (with its HU $> -500$ \cite{ct-lung}) has its intensity values similar to those of lesions, but it exhibits a regular pattern, which makes it amenable for modeling say by a deep network. This fact motivates us to formulate lesion segmentation as a \textbf{voxel-level anomaly modeling} problem. 

We hypothesize that if all the normal signals are captured at a voxel level, then the remaining abnormal voxels are localized automatically, which are grouped together as lesions. To facilitate voxel-level anomaly modeling, we design a novel proxy task. Firstly, we manually produce anomalies as synthetic `lesions' and insert them into normal CT images, forming pairs of normal and `abnormal' images for training. The `lesion' synthesis procedure constitutes a few simple operations, such as random shape generation, random noise generation within the shape and traditional filtering. Then using these training pairs, we learn a deep image-to-image network that recognizes normal textures from synthetic anomalies images. The state-of-the-art 3D image segmentation model, 3D U-Net\cite{3dunet}, is adopted as our deep network, which we call as a normalcy-recognizing network (NormNet). 

In practice, we increase the difficulty of the proxy task by building a `lesion' library \textit{as rich as possible}, which serves as a \textbf{superset} relative to the COVID-19 lesions. To distinguish normal contexts from these various anomalies, the NormNet is learned to be highly sensitive to the normal contexts, resulting in a tight decision boundary around the distribution of normal tissues. Finally, as shown in Fig.~\ref{fig:intro}(b), this boundary can also be used to segment COVID-19 lesions. We validate the effectiveness of NormNet on three different public datasets. Experimentally, it clearly outperforms various competing label-free approaches and its performances are even comparable to those of supervised method by some metrics. 


It should be noted that our approach differs from a research line called unsupervised anomaly detection (UAD) \cite{anomalysurvey09,anomalysurvey09med,mri-review,inpainting-anomaly}, which aims to detect the out-of-distribution (OOD) data by memorizing and integrating anomaly-free training data and has been successfully applied in many instance-level holistic classification scenarios. Further, our method differs from those methods in the inpainting \cite{inpainting} task, whose images in both training and testing sets are contaminated by the masks (noises) from the same domain. Finally, our method is different from synthetic data augmentation \cite{synthetic}, which manually generates lesions according to the features generated from labeled lesion area. In contrast, we do not need any image with labeled COVID-19 lesions.

In summary, we make the following contributions:
\begin{itemize}
    \item We propose the NormNet, a voxel-level anomaly modeling network, to distinguish healthy tissues from the COVID-19 lesion in the thorax area. 
    This training procedure only needs a large-scale healthy CT lung dataset, without any labeled COVID-19 lesions.
    \item We design an effective strategy for generating synthetic ‘lesions’ using only three simple operations: random shape, noise generation, and image filtering.
    \item The experiments show that our NormNet achieves better performances than various competing label-free methods on three different COVID-19 datasets. 
\end{itemize}

\section{Related Work}

\subsection{COVID-19 screening and segmentation for chest CT}

Deep learning based methods for chest CT greatly help COVID-19 diagnosis and evaluation~\cite{covid-review,ct-role}. Wang et al.~\cite{tmi-wang} proposed a weakly-supervised framework for COVID-19 classification at the beginning of the pandemic, which achieved high performance. Wang et al.~\cite{tmi-attention} exploited prior-attention residual learning for more discriminative COVID-19 diagnosis. Ouyang et al.~\cite{tmi-dual} solved the imbalanced problem of COVID-19 diagnosis by a dual-sampling attention network. However, it is more difficult for the COVID-19 segmentation task due to the lack of well-labeled data~\cite{tmi-inf}, lesion diversities~\cite{radioreport1} and noisy labels~\cite{tmi-noise}. Researchers have made attempts to address the above challenges. For example, to tackle the problem of labeled data scarcity, Ma et al.~\cite{junma} annotated 20 CT volumes from coronacases \cite{datasetseg3} and radiopedia \cite{datasetseg2}. Fan et al.~\cite{tmi-inf} proposed a semi-supervised framework called Inf-Net. Zhou et al.~\cite{tmi-rapid} solved the same issue by fitting the dynamic change of real patients’ data measured at different time points. However, all of these models depended on data with semantic labels. In this work, we propose an unsupervised anomaly modeling method called NormNet, which achieves comparable performances, but with no need of labeled data.

\begin{figure*}[t]
\centering
\centerline{\includegraphics[scale=0.5]{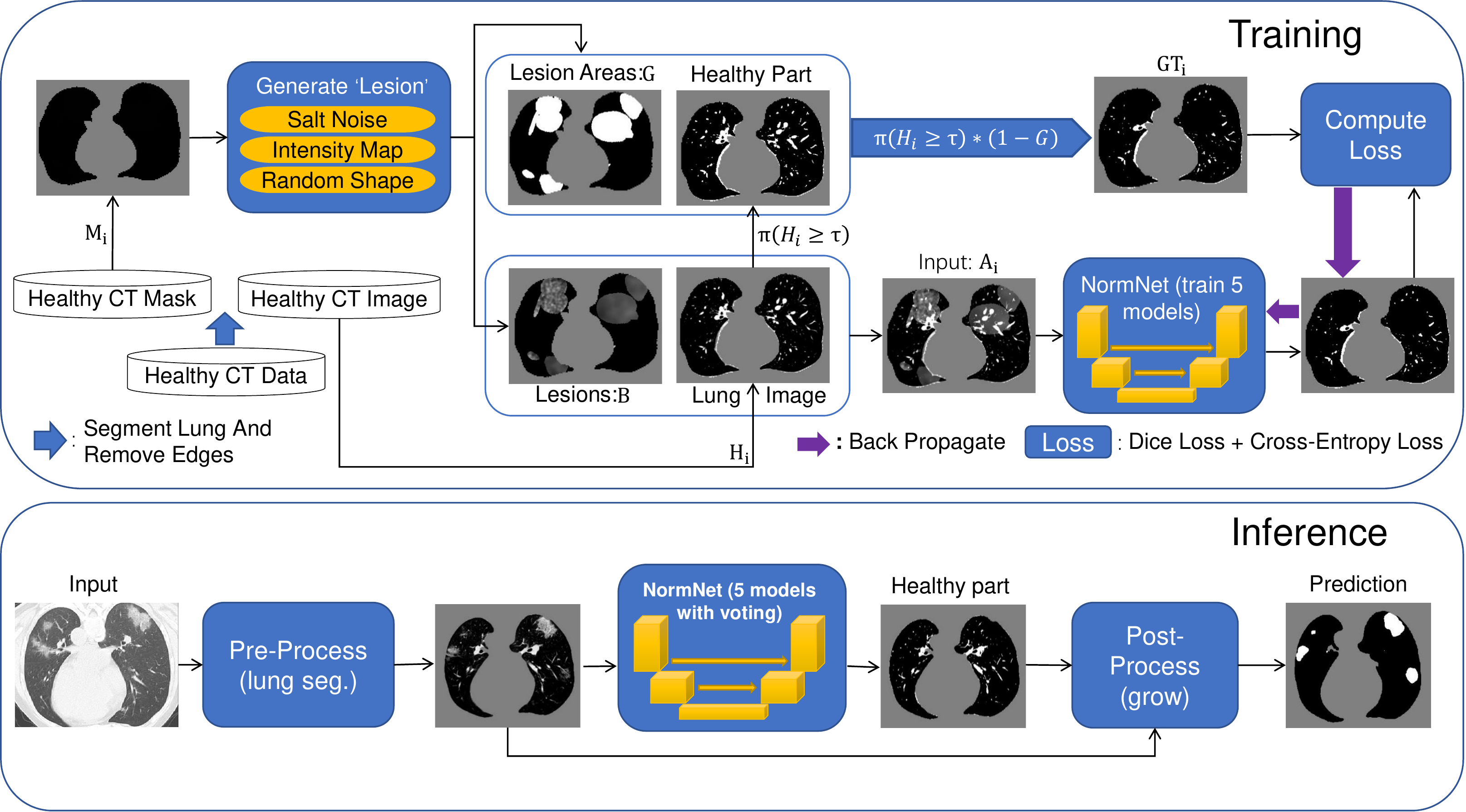}}
\caption{The overall framework of proposed NormNet. We first insert random noises (fake `lesions') $B$ to the healthy lung area $M_i$. Then we train the NormNet to segment the healthy tissues in the healthy area $M_i \odot (1-G)$ and high intensity range ($H_i > \tau$). In the inference time, the NormNet segment healthy tissues precisely and treat the remaining COVID-19 lesions as anomalies.
}
\vspace{-3mm}
\label{main}
\end{figure*}

\subsection{Anomaly detection}
\label{Sec:related-anomaly}

Anomaly detection or outlier detection is a lasting yet active research area in machine learning \cite{pangreview,pangkdd,pangcovid}, which is a key technique to overcome the data bottleneck \cite{vae-anomaly}. A natural choice for handling this problem is one-class classification methods, such as OC-SVM \cite{ocsvm}, SVDD \cite{svdd}, Deep SVDD \cite{dsvdd} and 1-NN. These methods detect anomaly by clustering a discriminate hyper-lane surrounding the normal samples in the embedding space.

In medical image analysis, there was another line of research which successfully detected anomaly in instance-level by finding the abnormal area \cite{OCT-TMI}. Recently, CNN-based generative models such as Generative Adversarial Networks (GAN) \cite{GAN}, and Variational Auto-encoders (VAE) \cite{VAE} have been proved essential for unsupervised anomaly segmentation \cite{f-anogan}. These methods first captured the normal distribution by learning a mapping between the normal data and a low-dimensional latent space by reconstruction loss. They assumed that if this process is only trained with normal distributions, a lesion area with abnormal shape and context can not be correctly mapped and reconstructed, resulting in high reconstruction error, which helped to localize the lesion area. The f-AnoGAN method \cite{anogan} learned the projection by solving an optimization problem, while VAE \cite{VAE} tackled the same problem by penalizing the evidence lower bound (ELBO). Several extensions such as context encoder \cite{context}, constrained VAE \cite{constrained}, adversarial autoencoder \cite{constrained}, GMVAE \cite{GMVAE}, Bayesian VAE \cite{Baysian} and anoVAEGAN \cite{spatial-anovaegan} improved the accuracy of the projection. Based on the pretrained projection, You et al. \cite{GMVAE} restored the lesion area by involving an optimization on the latent manifold, while Zimmerer et al. \cite{vae-anomaly} located the anomaly with a term derived from the Kullback-Leibler (KL)-divergence. 

Different from classification, lesion segmentation usually depends on locally fine-grained texture information. Unluckily, the decoder may loose some detailed texture information~\cite{SSIM}, which limited the accuracy of the reconstruction and caused false-positives. To make matters worse, as shown in Fig.~\ref{fig:intro}(a), healthy textures in Lung CT are fine-grained and need a more precise reconstruction. On the other hand, the calibrated likelihood of the decoder may not be precise enough \cite{cvpruniform}. The out-of-distribution data had some possibilities to be successfully reconstructed \cite{iccvmemory}, which raised false-negatives.


NormNet is designed to alleviate such issues by modeling the normal tiussue at a voxel level. Specifically, we propose a proxy task of separating healthy tissues from diverse synthetic anomalies. Firstly, we choose a 3D U-Net \cite{3dunet} as backbone, which uses the skip connection to alleviate the loss of information. Next, we make the appearance of synthetic `lesions' as diverse as possible to encourage our NormNet to be highly sensitive to normal textures. As a consequence, a tight decision boundary around normal tissues can be used to recognize healthy tissues and to segment COVID-19 lesions.  

\section{Method}
In this section, we firstly introduce the overall framework of our NormNet. Then we illustrate how to generate diverse `lesions'  in the given lung mask. Finally, we clarify how to post-process the healthy voxels predicted by our NormNet to obtain the final lesion mask for an unseen test image.

\subsection{Overall framework}

Let $\{R_1, R_2, \cdots, R_T\}$ be a set of $T$ healthy lung CT images. We clip the raw image $R_i$ with an HU range of $[-800,100]$ and scale the clipped image to $[0,1]$, obtaining $R'_i$. As shown in Fig.~\ref{main}, our method firstly use nnUnet\cite{nnunet} to obtain the lung masks $\{M'_1, M'_2, \cdots, M'_T\}$ and the thorax areas $\{H'_1, H'_2, \cdots, H'_T\}$ with $H'_i = R'_i \odot M'_i$, where $\odot$ stands for voxel-wise multiplication. It is worth noting that because no segmentation model can achieve $100\%$ accuracy, and there are always some edges caused by segmentation errors left in the thorax area $H'_i$, we introduce a simple pre-processing step (in Section~\ref{sec:edge}) to remove erroneous edges and generate a new lung mask $M_i$. Finally the thorax areas are updated to $\{H_1, H_2, \cdots, H_T\}$ with $H_i = H'_i \odot M_i$.



Then we use the synthetic `lesion' generator described in Section~\ref{sec:lesion} to synthesize various `lesions' $B$ within the lung masks $M_i$ with diverse shapes $G$ and textures, and inject them into the thorax area $H_i$ to form the input $A_i$.
Because the healthy voxels in the high-intensity range (say HU$\ge T$ with the threshold $T=-500$) have regular patterns and meaningful clinical content (tracheae and vessels~\cite{ct-lung}), we concentrate on segmenting normal patterns within high intensity range and normal areas. Accordingly, we compute ground truth as
\begin{align}
    GT_i = \pi(H_i \ge {\tau}) \odot (1 - G),
\label{Eq:gt}
\end{align}
where $\pi(.)$ is an indicator function that produces a binary mask. Note that the value of $\tau$ in $H_i$ is equivalent to the HU threshold; for example, $T=-500$ means $\tau =0.33$. 
Our NormNet is learned to predict the healthy part from $A_i$ via encouraging it to be close to $GT_i$ (aka minimizing Dice loss and cross-entropy loss). In this procedure, our NormNet learns to capture the context of healthy tissues quickly and precisely.

When our NormNet is applied to an unseen COVID-19 CT volume, it recognizes the healthy part of the volume with a high confidence and the lesion part of the volume with a low confidence. 
The confidence scores thus can be used as a decision boundary to predict the healthy parts and lesions. Because our training process is random, we form an ensemble by learning five random models under the same setting. A majority-vote for healthy parts is conducted as the final prediction. 

At last, we design a post-processing procedure in Section~\ref{sec:post} to obtain the final prediction.  As NormNet is trained to segment the voxels with HU$\ge T$, a small number of lesion voxels whose HU$<T$ are not taken into consideration and might get missed. So, we grow the localized lesion areas (in high-intensity range) to bring them back. 

\subsection{Removing erroneous edges} \label{sec:edge}

As mentioned above, this step is to separate the wrong edges caused by segmentation errors from lung mask $M'_i$. For a pair of inputs $\{M'_i, H'_i\}$, we select all the connected areas~\cite{component} in thorax area $H'_i$ with most of the voxels lying on the edges of the lung segmentation mask $M'_i$, and mark them as the wrong edges $E_i$. To avoid injecting noise into those edges, we use the lung mask without those edges, formulated as $M_i = M'_i \backslash E_i$. Note that we only launch this process in the training phase, leveraging the fact that no lesion occurs inside a healthy volume.

\subsection{Synthetic `lesion' generator} \label{sec:lesion}

\begin{figure*}[t]
\centerline{\includegraphics[scale=0.5]{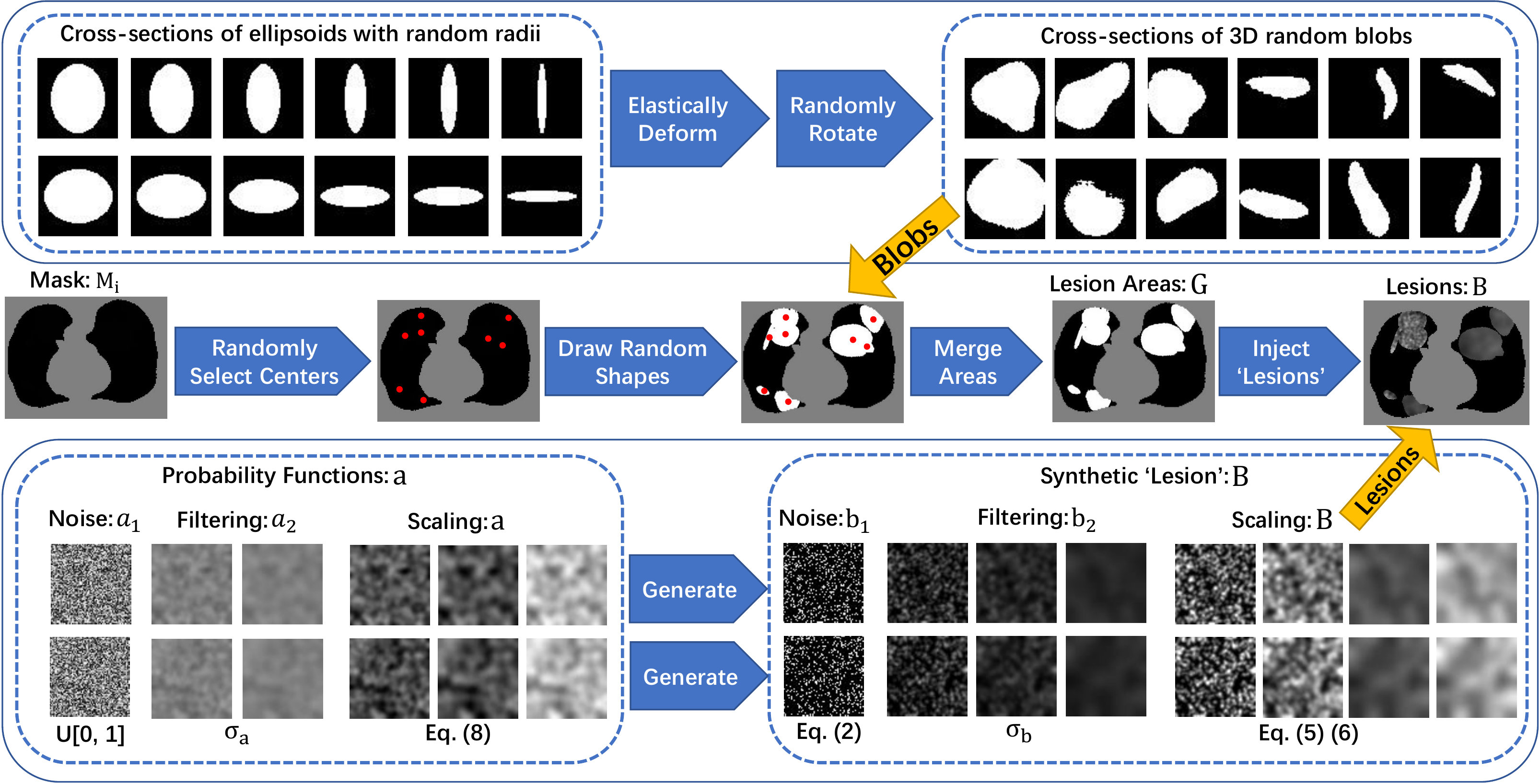}}
\caption{The schematic diagram of the proposed noise generator. We generate several diverse shapes and fill the connected areas with various filtered and scaled salt noises.}
\vspace{-3mm}
\label{noise}
\end{figure*}

As shown in Fig. \ref{noise}, the generator constitutes a set of simple operations, following the two steps: (i) generating lesion-like shapes; (ii) generating lesion-like textures. 
It is worth noting that all of the parameters are chosen for one purpose: \textit{generate diverse anomalies evenly.} The visualization of each step can be found in Supplementary Material. Below, we elaborate each step.

\subsubsection{Generating lesion-like shapes}\label{sec:shapes}

Multiple COVID-19 lesions may exist in a CT scan and they have various shapes. To obtain multiple lesion-like shapes with a CT, we propose the following pipeline. Below, $U[a,b]$ denotes a continuous uniform distribution within the range $[a,b]$, while $F[a,b]$ denotes a discrete uniform distribution.

\begin{itemize}
    \item 
For each lung mask $M_i$ with a shape of size $[32, 512, 512]$, compute a factor $\lambda=\frac{|M_i|}{|M_{max}|}$ to make sure that smaller masks generate fewer ellipsoids, where $M_{max}$ is the biggest mask in training set.

\item Create several ellipsoids as follows: (1) Sample a number $N_s \sim F[5\lambda,10\lambda]$ and then generate $N_s$ small-size ellipsoids with the principal semi-axes of each ellipsoid randomly selected from $U[3,10]$; (2)  Sample a number $N_m \sim F[5\lambda,10\lambda]$ and then generate $N_m$ medium-size ellipsoids with the principal semi-axes of each ellipsoid randomly selected from $U[10,32]$; and (3) Generate a large size ellipsoid with a probability of $P_L = 0.2\lambda$ and with its principal semi-axes 
$\sim U[32,64]$.

\item For each generated ellipsoid, deform it using elastic transformation \cite{elasticdeform} with random parameters and rotate it to align with random axes, yielding a blob $C$. Then position this blob at a random center inside the lung $H_i$. 
\end{itemize}
At this stage, we have a set of blobs $\{C_1, C_2, \ldots \}$. 
Then we merge connected blobs and obtain several non-adjacent blobs $\{G_1, G_2, \ldots \}$ with varying shapes. For each blob $G_j$, we synthesize a patch of ‘lesion’ $B_j$ by the following steps.


\subsubsection{Generating lesion-like textures}

The texture pattern of lesions varies\footnote{The only prior knowledge we used is that water, tissues, infections have much higher intensities than air in lung CT~\cite{ct-lung}.}; thus it is challenging to generate lesion-like textures. Below we outline our attempt of doing so using a set of simple operations. It should be noted that our method still has room for optimization, but it is already empirically effective.

We follow a series of three steps, namely noise generation, filtering~\cite{meanfiltering}, and scaling/clipping operations, to generate the lesion-like textures.
\begin{itemize}
    \item  Noise generation. For each voxel denoted by $x$, generate salt noise $b_1(x)$
\begin{align}
    b_1(x) = \big \{   
    \begin{tabular}{l}
        $1$~~\text{with a probability}~~$a(x)$;\\
        $0$~~\text{with a probability}~$1-a(x)$,
  \end{tabular}
\end{align}
where the voxel-dependent probability function $a(x)$ will be defined later.

\item Filtering~\cite{meanfiltering}. Filter the noise image $b_1(x)$ to obtain $b_2(x)$ using a Gaussian filter $g$ with a standard deviation $\sigma_b$.
\begin{align}
    b_2(x) = g(x; \sigma_b) \otimes b_1(x),  
\end{align}
where $\otimes$ is the standard image filtering operator.
The standard deviation $\sigma_b$ is randomly sampled as follows:
\begin{align}
    \sigma_b \sim  \big \{ 
    \begin{tabular}{l}
        $U[0.8,2]$~~\textrm{with a probability of 0.7};\\
        $U[~2,5]$~~\textrm{with a probability of 0.3}.
  \end{tabular}
\label{Eq:sigma_b}
\end{align}
.

\item Scaling and clipping. This yields the lesion-like pattern $B_j(x)$.
\begin{align}
    B_j(x) = clip_{[0,1]}( \beta b_2(x)),
\end{align}
with $\beta$ being the scaling factor that is obtained by 
\begin{align}
    \beta = \mu_0 / mean_{0.2}( b_2(x) ) ,
\label{Eq:Salt}
\end{align}
where $\mu_0 \sim U[0.4, 0.8]$ and 
$mean_{t}( f(x) )$ is the mean intensity of the image $f(x)$ that passes the threshold $t$.
\end{itemize}
Now, we describe how to obtain the voxel-dependent probability function $a(x)$, again using a series of noise generation, filtering~\cite{meanfiltering}, and scaling operations.
\begin{itemize}
    \item Noise generation. For each voxel $x$, independently sample the uniform probability $U[0,1]$ to get a noise image $a_1(x) \sim U[0,1]$.
    
\item Filtering. Filter the noise image $a_1(x)$ to obtain $a_2(x)$ using a Gaussian filter $g$ with a standard deviation $\sigma_a$.
\begin{align}
    a_2(x) = g(x; \sigma_a) \otimes a_1(x),  
\end{align}
where the standard deviation $\sigma_a \sim U[3,15]$.

\item Scaling.  This yields the desired function $a(x)$.
\begin{align}
    a(x) &= scale_{[a_{L}, a_{U}]}(a_2(x) ) \nonumber \\
    &= (a_{U}-a_{L}) * \frac{a_2(x) - a_{2,min}} { a_{2,max} - a_{2,min}} + a_{L},
\label{Eq:probablity}
\end{align}
where $a_{U} \sim U[0,0.3]$, $a_{L} \sim U[0,0.3]$ and $a_{U}-a_{L} > 0.15$.
\end{itemize}

Finally, we inject the synthetic lesions $B_j$ into the various blobs $G_j$, and place these blobs at random centers inside the lung area $H_i$. Mathematically, the image $A_i$ with synthetic `lesions' is generated by finding the maximum value of the lung area  $H_i$ and the synthetic lesions $B_j$ at each voxel point: \begin{align}
    A_i = \max(H_i, B_1, B_2, \cdots).
\end{align}
Our goal is to learn a network that takes $A_i$ as input and outputs $GT_i$.












\begin{figure}[htbp]
\centering
\includegraphics[width=0.8\columnwidth]{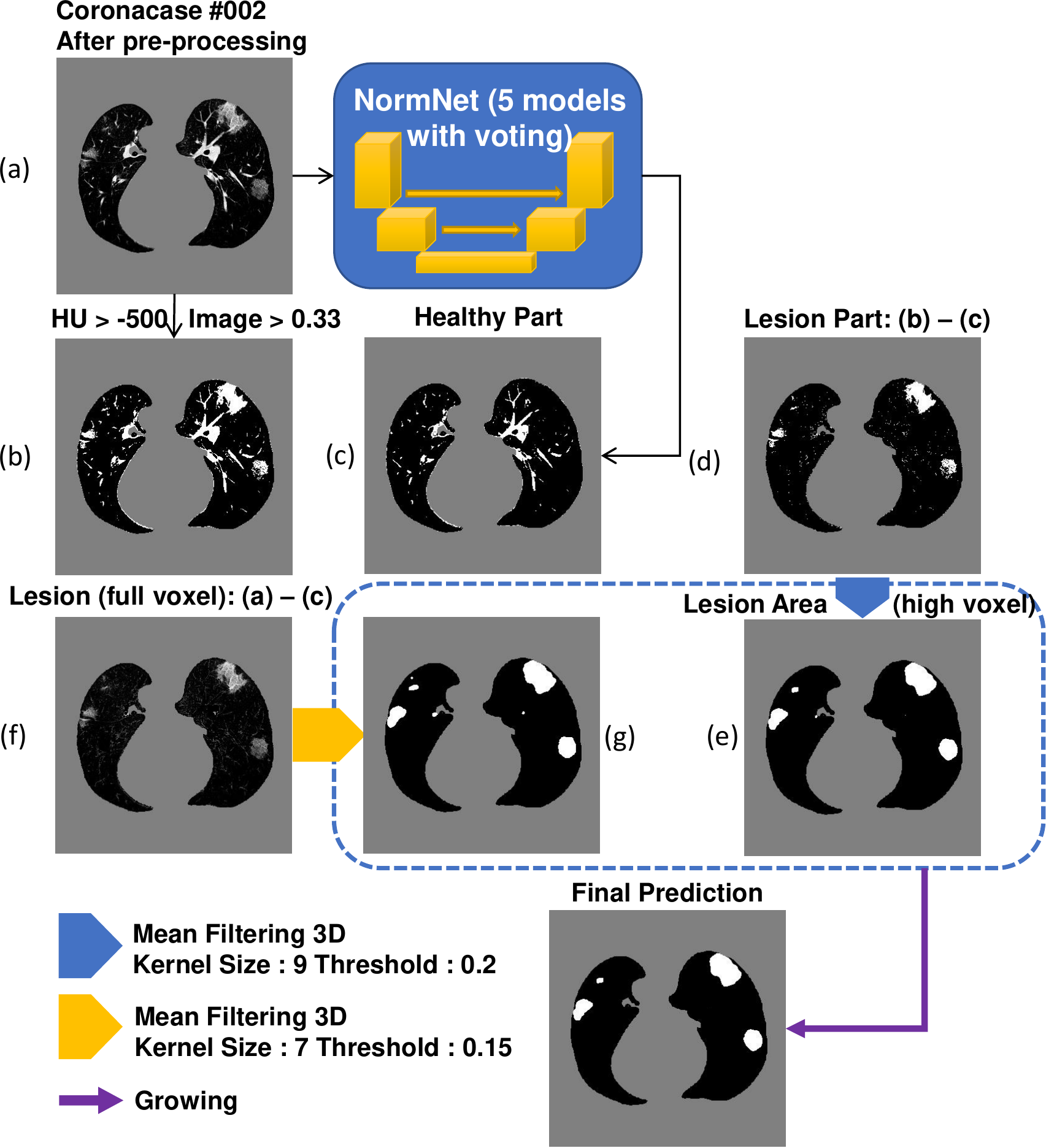}
\caption{The illustration of the post-processing process.This step removes the healthy part from the COVID-19 CT volume and generate final prediction by mean filtering and growing.}
\vspace{-3mm}
\label{post}
\end{figure}

\subsection{Post processing} \label{sec:post}

A post processing procedure is designed to obtain the final lesion prediction based on difference between the original CT volume and predicted healthy areas. As illustrated in Fig. \ref{post}, the final prediction is obtained with the following steps:

\begin{itemize}
\item Compute the lung mask (Fig. \ref{post}(b)) and predict the healthy part by NormNet (Fig. \ref{post}(c));

\item 
Compute the lesion region by subtracting the predicted healthy part from lung mask to get Fig. \ref{post}(d). Considering that only bright voxels $\ge {\tau}$ are in the lung mask, the full-voxel raw lesion areas (Fig. \ref{post}(f)) is calculated, aiming to `recover' less bright lesions;

\item Mean filtering $F$ with kernel size $k$ is then applied to Figs. \ref{post}(d) and \ref{post}(f) to smooth the lesion region (with kernel sizes of $k_d$ and $k_f$) and then to remove the background noise via thresholding (with thresholds of $t_d$ and $t_f$), 
which yields the results in Fig. \ref{post}(e) and \ref{post}(g), respectively;

\item Binary dilation~\cite{dilation} is used to grow the lesion regions of Fig.~\ref{post}(e) to bring the missing voxels in the low intensity range (HU $<$ T) back. Then, we remove the voxels out of the full lesion regions defined by Fig.~\ref{post}(g) to prevent over-growing. 
\begin{align}
  I_{final} = Dilation(I_e) * I_g,
\label{Eq:Dialation}
\end{align}
where $I_{final}$ is the prediction in Fig.~\ref{post}; $I_e$ is Fig.~\ref{post}(e), and $I_g$ is Fig.~\ref{post}(g).

\end{itemize}

\section{Experiments}

Below we firstly provide a brief description of the various CT lung datasets used in our experiments. 
Then we present our experimental settings and the baseline approaches we implement and compare. Finally, we show our main experimental results, hyper-parameter analyses and an ablation study. 

\subsection{Datasets}

One distinguishing feature of the paper lies in unleashing the power embedded in existing datasets. Rather than using a single dataset, we seamlessly integrate multiple CT lung datasets for three different tasks of healthy lung modeling, COVID-19 lesion segmentation, and general-purpose lung segmentation into one working solution.

\subsubsection{CT datasets for healthy lung modeling}

LUNA16 \cite{luna16} is a grand-challenge on lung nodule analysis. The images are collected from The Lung Image Database Consortium image collection (LIDC-IDRI) \cite{NSCLC3,LIDC1,LIDC2}, and each image is labeled by 4 experienced radiologists. As half of the images are healthy and clean except for those containing nodule areas, we select 453 CT volumes from LUNA16 and remove the slices with nodules to formulate our healthy lung CT dataset.

\subsubsection{CT datasets for COVID-19 lesion segmentation }

To measure the performance of our methods towards COVID-19 segmentation, we choose two public COVID-19 CT segmentation datasets in  Table \ref{public} and one UESTC with semantic labels. It is worth noting that our method segments the COVID-19 lesions under the unsupervised 
setting, and thus the labeled datasets are only used for testing.

\begin{itemize}
    \item \emph{Coronacases:} There are 10 public CT volumes in the \cite{datasetseg3} uploaded from the patients diagnosed with COVID-19. These volumes are firstly delineated by junior annotators\footnote{Ma et al. provide 20 well-labeled CT volumes, in addition to the 10 volumes of coronacases, the other 10 volumes have been clipped to [0 -- 255] without any information about HU, which is not applicable based on our methods.}, and then refined by two radiologists with 5 years experience, and finally, all the annotations are verified and refined by a senior radiologist with more than 10 years experience in chest radiology diagnosis \cite{junma}.
    \item \emph{Radiopedia:} Another 8 axial volumetric CTs are released from Radiopaedia \cite{datasetseg2} and have been evaluated by a radiologist as positive with voxel-wise labeling on lesion regions \cite{datasetseg1}.
    \item \emph{UESTC:} A large-scale well labeled datasets~\cite{uestc} containing 120 CT volumes, of which 50 are labeled by experts and 70 by non-experts.
\end{itemize}

\subsubsection{CT datasets for general purpose lung segmentation }

To obtain the accurate lung area in the CT volume, we choose nnU-Net \cite{nnunet} as our lung segmentation method, which is proved to be state-of-the-art segmentation framework in medical imaging analysis. We use three lung CT datasets with semantic labels for the lung region:

\begin{itemize}
    \item \emph{NSCLC left and right lung segmentation:} This dataset consists of lung volume segmentation collected on 402 CT scans from The Cancer Imaging Archive NSCLC Radiomics \cite{NSCLC1,NSCLC2,NSCLC3}.
    \item \emph{StructSeg lung organ segmentation:} This dataset consists of 50 lung cancer patient CT scans with lung organ segmentation. The dataset served as a segmentation challenge during MICCAI 2019 \cite{structseg}.
    \item \emph{MSD Lung tumor segmentation} This dataset consists of 63 labelled CT scans, which served as a segmentation challenge during MICCAI 2018 \cite{msd}. The lung regions are labeled by Ma et al. \cite{junma}.
\end{itemize}

We choose 2D U-Net as the backbone. The model is trained by nnU-Net \cite{nnunet} in 5-fold cross-validation, which segments the lung region very precisely with Dice scores larger than 0.98 in both Coronacases and Radiopedia datasets.

\subsection{Experimental settings}

\subsubsection{Evaluation metrics}

We use several metrics widely used to measure the performance of segmentation models in medical imaging analysis, including precision score (PSC), sensitivity (SEN) and Dice coefficient (DSC), which are formulated as follows:
\begin{align}
PSC=\frac{tp}{tp+fp}; SEN= \frac{tp}{tp+fn} ; DSC=\frac{2tp}{2tp+fn+fp}, \nonumber
\end{align}
where $tp$, $fp$ and $fn$ refer to the true positive, false positive and false negative respectively.

\subsubsection{Pre-processing}
All of the images in the training and testing sets are segmented for the lung region at first. Then we unify their spacing to $0.8 \times 0.8 \times 1 mm^3$, as well as orientation. Next, all of the images are clipped with window range $[-800, 100]$ and normalized to $[0, 1]$. Finally, the lung regions are centralized and padded to $512\times512$ with 0.

\begin{figure*}[t]
\centering
\centerline{\includegraphics[width=2\columnwidth]{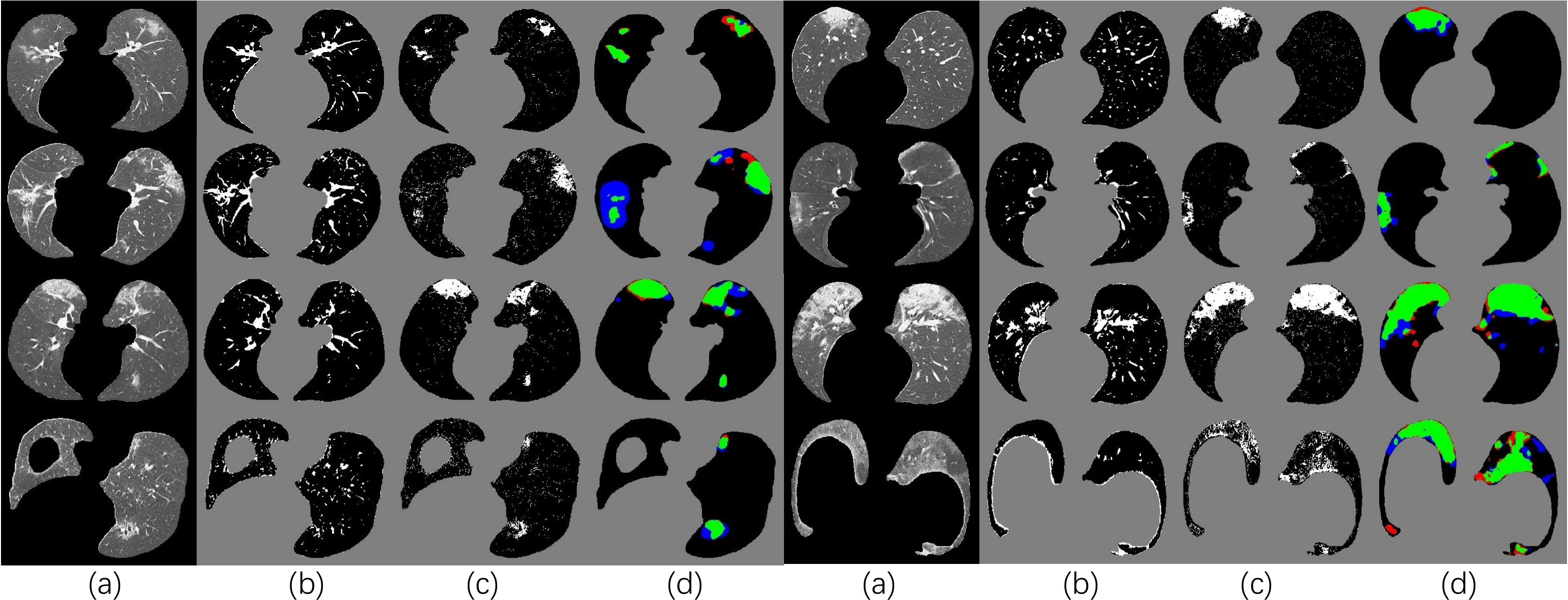}}
\caption{Visual results of our NormNet for COVID-19 segmentation. (a), (b), (c) and (d) represents input (after lung segmentation), healthy tissues (predicted from our NormNet), lesion parts, and final segmentation, respectively. The green, blue, and red areas in (d) refer to true positive, false negative, and false positive, respectively.}
\vspace{-3mm}
\label{results}
\end{figure*}

\subsubsection{Training and inference details}
We choose 3D U-Net \cite{3dunet} as backbone for NormNet, implemented by MONAI \cite{MONAI}. As all of the volumes in both training and testing phases are well aligned, no more augmentation is needed. The NormNet is trained on a TITAN RTX GPU and optimized by the Adam optimizer~\cite{Adam} with default settings. We train our network for 2000 iterations with a batch size of 8, and set the learning rate to 3e-4.
For the testing phase, as the contexts of healthy signals are precisely captured by our NormNet, these signals are predicted with high probability. Therefore, we select those voxels with probability $> 0.95$ as healthy parts in the COVID-19 CT volume. For the mean filtering in the post processing, we set kernel sizes ($k_d, k_f$) to (9, 7) and thresholds ($t_d, t_f$) to (0.2, 0.15) \footnote{We use hyper-parameter search for the 4 parameters on 3 CT volumes from the dataset `Coronacase'} for lesion parts with bright voxels (Fig. \ref{post}d) and full voxels (Fig. \ref{post}f), respectively. We obtain these values according to the hyperparameter search, which are fixed to all of three COVID-19 datasets.

\subsection{Baselines}

We compare our methods with existing deep learning based methods\footnote{We follow the experimental setting in~\cite{mri-review} and use their source code: \url{https://github.com/StefanDenn3r/unsupervised\_anomaly\_detection\_brain\_mri}.} in medical imaging analysis for unsupervised anomaly detection (UAD) methods to evaluate the effectiveness of our approach. To eliminate the influence of irrelevant factors, we use the images with only lung regions as training and testing sets for all of the experiments (except for \textbf{VAE Original}). These encoder-decoder based methods are trained with a learning rate of 3e-4 and a batch size of 16 for 6000 iterations. To obtain the best performance for each method, we perform a greedy search up to two decimals to get the threshold with best Dice score for each COVID-19 dataset.

\begin{itemize}
    \item \textbf{AE:} An Autoencoder with a dense bottleneck to learn a mapping between latent space $z\in\mathbb{R}^{128}$ and input space $\mathbb{R}^{D\times H \times W}$, which assumes that only normal input can be successfully reconstructed.
    \item \textbf{VAE \cite{VAE}:} Different to \textbf{AE}, VAE use KL-divergence and resampling to constrain the latent space. As the reconstruction is more difficult for lung CT images, so we set $\alpha$ for KL loss as 1e-6.
    \item \textbf{VAE Spatial \cite{spatial-anovaegan}:} A \textbf{VAE} with a spatial (fully-convolutional) bottleneck, which learns a mapping between latent space $z\in\mathbb{R}^{8\times8\times128}$ and $\mathbb{R}^{D\times H \times W}$.
    \item \textbf{VAE Original:}  A \textbf{VAE} trained with the full lung CT images, instead of lung regions (after lung segmentation).
    \item \textbf{Context VAE \cite{context}:} An expansion of \textbf{VAE}, which forces the encoder of \textbf{VAE} to capture more information by reconstructing a input image with cropped patches.
    \item \textbf{Constrained VAE \cite{constrained}:} An expansion of \textbf{VAE}, which uses the encoder to map the reconstructed image to the same point as the input in the latent space.
    \item \textbf{GMVAE \cite{GMVAE}:} An expansion of \textbf{VAE}, which replaces the mono-modal prior of the VAE with a Gaussian mixture.
    \item \textbf{Bayesian VAE \cite{Baysian}:} An expansion of \textbf{VAE}, which aggregates a consensus reconstruction by Monte-Carlo dropout.
    \item \textbf{KL Grad \cite{vae-anomaly}:} Use the gradient map derived from KL loss to segment anomalies.
    \item \textbf{VAE restoration \cite{GMVAE}:} Restore the abnormal input to decrease the evidence lower bound (ELBO). The restoration part is marked as the detected abnormal area.
    \item \textbf{f-AnoGAN \cite{f-anogan}:} Different from \textbf{VAE}, f-AnoGAN learns such a mapping by solving an optimization problem. To keep the training process of f-Anogan stable, we resize the lung image to $[64,64]$ after center crop. 
\end{itemize}

In order to reveal the top-line for each dataset, we train \textbf{nnU-Net} \cite{nnunet} in 5-fold cross-validation\footnote{80\% of the data are used in the training set.}. Furthermore, to test the performance of the supervised model when inferring unseen datasets, we train nnU-Net on two COVID-19 datasets and test on the remaining one, called \textbf{nnU-Net-Unseen}. At last, we test the pretrained models\footnote{Released in their official websites. COPLE-Net: \url{https://github.com/HiLab-git/COPLE-Net}; Inf-Net: \url{https://github.com/DengPingFan/Inf-Net}} of two (semi-)supervised methods of COPLE-Net~\cite{tmi-noise} and Inf-Net~\cite{tmi-inf} on the same datasets.


\subsection{Segmentation results}

\begin{figure}[t]
\centering
\centerline{\includegraphics[width=0.9\columnwidth]{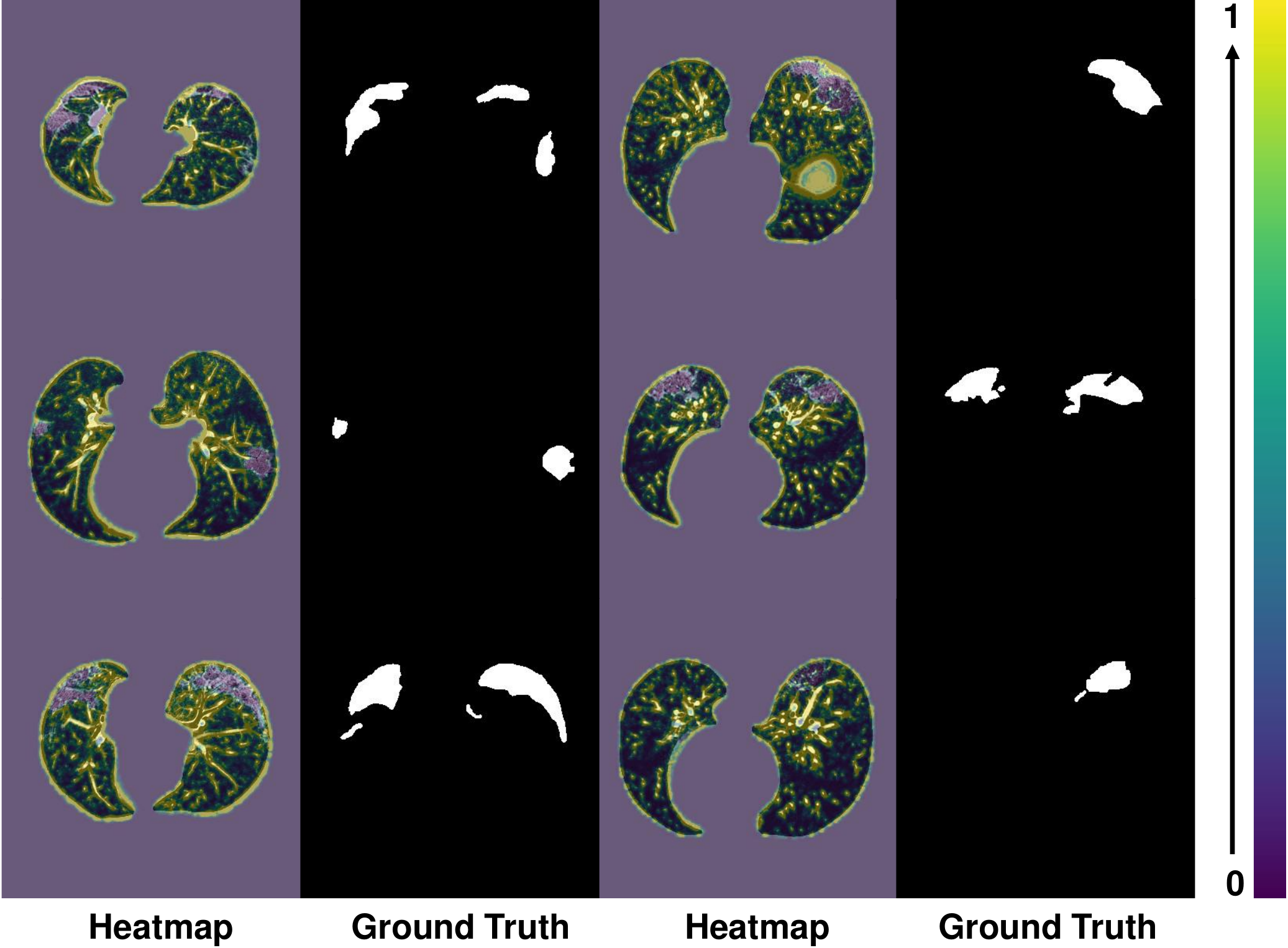}}
\caption{The heatmap from the first down-sampling block of the NormNet. The NormNet captures the contexts of healthy tissues precisely.}
\vspace{-3mm}
\label{Fig:Heatmap}
\end{figure}

As shown in Fig.~\ref{Fig:Heatmap}, NormNet is much more sensitive to the contexts of healthy voxels than possible anomalies (COVID-19 lesions our work). To validate the ability of our NormNet to recognize healthy voxels from anomalies, we collect all bright voxels with $\tau \ge 0.33$ of the CT volumes. As in Table \ref{highvoxel}, our method successfully recognizes healthy voxels from the COVID-19 lesion voxels with AUC larger than 85\%. The high specificity ensures that most of the lesions are treated as anomaly. Our NormNet firstly votes for the healthy tissues from the CT volumes with COVID-19 lesions. Then, the post-processing procedure grows the lesion area to contain more lesions with less bright voxels ($\tau < 0.33)$. We also use mean filtering in the post-processing to remove the isolated healthy voxels that are segmented as anomaly, as shown in Fig. \ref{results}{c}. 

Therefore, our method reaches the Dice scores of 69.8\%, 59.3\%\footnote{
We remove CT volume \#6 from the Radiopedia dataset as it has only about 70 positive voxels in 42 slices.} 
and 61.4\%\footnote{We select the CT volumes with the spacing of z-axis less than 5 mm, since our training set (LUNA16) only contains CT volumes with the spacing of z-axis less than 5 mm.}  (shown in Table \ref{dice}) in the three different COVID-19 datasets respectively.
The visual results shown in Fig. \ref{results} reveal that most of the COVID-19 lesions are successfully (green area) segmented by our NormNet. Furthermore, without the expensive annotations, our NormNet achieves competitive performances on the three public datasets against these (semi-)supervised models. The performance of Inf-Net~\cite{tmi-inf}, which is overall similar to that of our NormNet,  is rather stable across the three datasets as it is trained based on a different dataset. However, on the largest dataset (UESTC), NormNet still has performance gaps (the Dice scores of 10.2\%, 17.4\% and 20.2\%), compared to the supervised methods. Specifically NormNet has precision gaps on Radiopedia (18.7\%) and UESTC (21.7\%), as well as sensitivity gaps on Coronacases (16.1\%) and Radiopedia (14.9\%).


\begin{table}[htbp]
\centering
 \caption{The results of segmentation performances of bright voxels.}
 \begin{tabular}{lrrrr} 
  \toprule 
Dataset & Precision & Sensitivity & Specificity  & AUC    \\ 
  \midrule 
Coronacases &  90.5     &   78.6    & 81.2   & 87.0    \\
Radiopedia           &93.1   & 70.9   &86.9  &89.7     \\
UESTC      & 92.1   & 77.6   &84.7  &  88.4 \\
  \bottomrule 
 \end{tabular} 
\label{highvoxel}
\end{table}

\begin{table*}[htb]
 \caption{The quantitative results of our method compared to other UAD methods and nnU-net. For each column, the \textbf{top}, \textcolor{red}{second} and \textcolor{blue}{third} values are highlighted.} 
 \centering
 \begin{tabular}{l|rrr|rrr|rrr} 
  \toprule 
 \multirow{2}*{Methods} & \multicolumn{3}{|c}{Coronacases}  & \multicolumn{3}{|c}{Radiopedia}  & \multicolumn{3}{|c}{UESTC} \\
  & DSC (\%) & PSC (\%) & SEN (\%) &  DSC (\%) & PSC (\%) & SEN (\%) & DSC (\%) & PSC (\%) & SEN (\%)    \\
  \midrule 
  nnU-Net \cite{nnunet} &80.1$\pm$6.73 &80.2$\pm$12.4 &82.3$\pm$9.30 &76.7$\pm$5.81 & 77.1$\pm$14.0 &80.5$\pm$13.1 &81.6$\pm$9.43 & 83.0$\pm$11.7 &81.7$\pm$11.6  \\  
 nnU-Net-Unseen &78.1$\pm$12.0 &79.0$\pm$13.6 &78.9$\pm$13.5 &68.0$\pm$19.8 &60.9$\pm$22.9 &87.6$\pm$9.72 &79.3$\pm$8.19 & 82.8$\pm$12.8 &77.6$\pm$8.71  \\  
 COPLE-Net~\cite{tmi-noise} & 68.2$\pm$10.8 & 77.3$\pm$9.57 & 63.4$\pm$16.9 &59.3$\pm$17.7 & 57.9$\pm$16.0 & 63.3$\pm$23.0 &83.9$\pm$9.47 & 84.8$\pm$10.5 & 84.6$\pm$12.0 \\
 Inf-Net~\cite{tmi-inf} & 66.9$\pm$14.0 & 74.5$\pm$15.6 & 63.3$\pm$16.5 & 67.8$\pm$13.3 & 65.2$\pm$17.1 & 75.9$\pm$13.4 &63.9$\pm$11.8 & 70.4$\pm$17.9 & 62.2$\pm$12.8 \\
 
  \midrule
AE &28.3$\pm$15.5	&21.5$\pm$15.3	&52.1$\pm$11.3	&30.3$\pm$17.7	&24.4$\pm$19.0	&58.9$\pm$6.2 &21.1$\pm$22.5 &\textcolor{blue}{21.9$\pm$25.7} &44.0$\pm$11.1    \\
VAE \cite{VAE}   &26.4$\pm$14.5	&19.8$\pm$14.0	&50.1$\pm$9.8	&28.1$\pm$17.5	&21.6$\pm$17.6	&\textcolor{blue}{62.3$\pm$5.7} &21.3$\pm$20.4 &16.7$\pm$20.1 &44.6$\pm$11.7	   \\
VAE Spatial \cite{spatial-anovaegan} &27.4$\pm$16.5	&21.0$\pm$16.4	&49.9$\pm$11.9	&30.7$\pm$19.8	&\textcolor{blue}{24.8$\pm$20.7}	&59.2$\pm$8.0 &25.4$\pm$22.2 &18.3$\pm$26.4 &42.7$\pm$13.5	   \\
VAE Original &10.9$\pm$8.0	&6.9$\pm$6.1	&41.3$\pm$8.2	&12.3$\pm$10.5	&8.5$\pm$8.9	&44.9$\pm$4.9 &10.2$\pm$10.4 &5.2$\pm$11.6 &31.9$\pm$4.7	  \\
Context VAE \cite{context}  &\textcolor{blue}{29.7$\pm$16.0}	&\textcolor{blue}{21.8$\pm$15.6}	&\textcolor{blue}{61.0$\pm$9.8}	&\textcolor{blue}{32.3$\pm$21.3}	&24.3$\pm$20.6	&72.2$\pm$6.0 &\textcolor{blue}{27.2$\pm$26.7} &19.2$\pm$25.8 &52.2$\pm$9.4	    \\
Constrained VAE \cite{constrained}   	&27.9$\pm$14.8	&21.0$\pm$14.7	&53.2$\pm$10.5	&29.2$\pm$17.7	&22.9$\pm$18.3	&61.3$\pm$5.6 &22.2$\pm$17.3 &17.8$\pm$20.4 &39.3$\pm$7.4	    \\
GMVAE \cite{GMVAE} &25.7$\pm$16.4	&20.2$\pm$14.4	&51.0$\pm$12.6	&28.6$\pm$17.7	&22.3$\pm$19.5	&\textcolor{red}{63.3$\pm$7.2} &24.7$\pm$20.3 &18.8$\pm$26.4 &40.9$\pm$11.8	  \\
Bayesian VAE \cite{Baysian} 	&27.5$\pm$15.0	&20.8$\pm$14.7	&50.9$\pm$11.4	&29.6$\pm$16.8	&23.5$\pm$17.6	&58.2$\pm$6.8 &22.0$\pm$17.2 &15.7$\pm$16.0 &40.1$\pm$12.3	  \\
KL Grad \cite{vae-anomaly} &9.5$\pm$8.2 &5.5$\pm$5.2 &\textcolor{red}{65.5$\pm$19.7} &10.2$\pm$14.2 &6.7$\pm$10.3 &39.1$\pm$20.3 &7.9$\pm$12.8 &6.2$\pm$9.4 &\textcolor{blue}{56.7$\pm$19.5} \\
VAE Restoration \cite{GMVAE} &12.8$\pm$4.5 &16.3$\pm$10.1 &12.1$\pm$2.5 &9.1$\pm$3.7 &16.5$\pm$16.0 &8.8$\pm$1.6 &6.4$\pm$2.8 &13.1$\pm$14.3 &7.0$\pm$1.0  \\
f-AnoGAN \cite{f-anogan} &15.4$\pm$12.6 &10.8$\pm$10.8 &38.3$\pm$13.2 &19.7$\pm$17.3 &14.2$\pm$14.9 &55.2$\pm$8.9 &12.1$\pm$11.8 &8.7$\pm$13.1 &40.3$\pm$8.2 \\

  \midrule
  Proposed w/o growing &\textcolor{red}{67.1$\pm$17.7} &\textbf{85.7$\pm$6.89} &60.0$\pm$22.5 &\textcolor{red}{54.6$\pm$17.4} &\textbf{59.2$\pm$18.6} &54.4$\pm$17.7 &\textbf{61.5$\pm$18.3} &\textbf{68.1$\pm$25.2} &\textcolor{red}{69.1$\pm$21.1} \\
  Proposed   &\textbf{69.8$\pm$15.2} &\textcolor{red}{82.1$\pm$8.92} &\textbf{66.2$\pm$22.2} &\textbf{59.3$\pm$16.9} & \textcolor{red}{58.3$\pm$18.0} &\textbf{65.6$\pm$18.7} &\textcolor{red}{61.4$\pm$19.4} &\textcolor{red}{61.3$\pm$26.1} &\textbf{77.6$\pm$19.6} \\
  
  \bottomrule 
 \end{tabular} 
\label{dice}
\end{table*}

\begin{table*}
 \caption{The ablation study for modules of `lesion' generator, threshold $T$,  hyper-parameters of `lesion' generator and post-processing. The dice score is used as metrics.} 
 \centering
 \begin{tabular}{l|rrrrrrrrrrrrrr|r} 
  \toprule 
\multirow{2}{*}{Dataset} & \multicolumn{14}{c|}{Hyper-parameters of `lesion' generator} & \multirow{2}{*}{NormNet}  \\
& \romannumeral1 & \romannumeral2 & \romannumeral3 & \romannumeral4 & \romannumeral5 & \romannumeral6 & \romannumeral7 & \romannumeral8 & \romannumeral9 & \romannumeral10 & \romannumeral11 & \romannumeral12 & \romannumeral13 & \romannumeral14 &  \\
   \midrule 
Coronacases & 68.8 & 68.0 & 70.5 & 70.8 & 69.1 & 69.5 & 68.5 & 66.5 & 66.9 & 70.6 & 66.2 & 68.2 & 66.5 & 64.1 & 69.8\\
Radiopedia & 60.8 & 61.0 & 60.4 & 55.0 & 57.0 & 58.9 & 57.3 & 59.1 & 60.7 & 58.4 & 58.3 & 60.3 & 54.6 & 53.2 & 59.3\\
UESTC & 61.2 & 61.3 & 61.6 & 60.3 & 63.0 & 61.4 & 60.7 & 60.9 & 62.2 & 62.2 & 60.5 & 62.7 & 59.9 & 60.8 & 61.4\\
\bottomrule
 \end{tabular} 

\vspace{2mm}
 
\begin{tabular}{l|rrrr|rrrr|rrr|rrr|r}
\toprule
 \multirow{2}{*}{Dataset} &  \multicolumn{4}{c|}{\textit{Param.} of post-processing} & \multicolumn{4}{c|}{Threashold $T$} & \multicolumn{3}{c|}{Modules of generator} & \multicolumn{3}{c|}{Other strategies} & \multirow{2}{*}{NormNet} \\
 
& $k_d$ & $k_f$ & $t_d$ & $t_f$ & -700 & -600 & -400 & -300 & $B_i$ & $a_i$ & $G_i$ & Edge & Areas & Lesions   \\
 
    \midrule 
Coronacases & 70.0 & 69.8 & 70.1 & 67.1 & 57.6 & 67.3 & 69.7 & 60.7 & 37.9 & 64.5 & 51.9 & 68.7 & 45.3 & 55.2 & 69.8 \\
Radiopedia & 59.7 & 60.0 & 60.0 & 58.9 & 52.2 & 59.7& 55.7 & 54.0 & 40.8 & 55.6 & 55.1 & 52.6 & 44.5 & 55.3 & 59.3\\
UESTC & 61.0 & 61.2 & 60.7 & 61.4 & 51.9 & 62.4 & 60.2 & 54.9 & 38.4 & 56.9 & 47.3 & 57.1 & 35.2 & 55.4 & 61.4\\
 
  \bottomrule 
 \end{tabular} 
\label{Table:ablation}
\end{table*}

\begin{figure}[t]
\centering
\includegraphics[width=0.8\columnwidth]{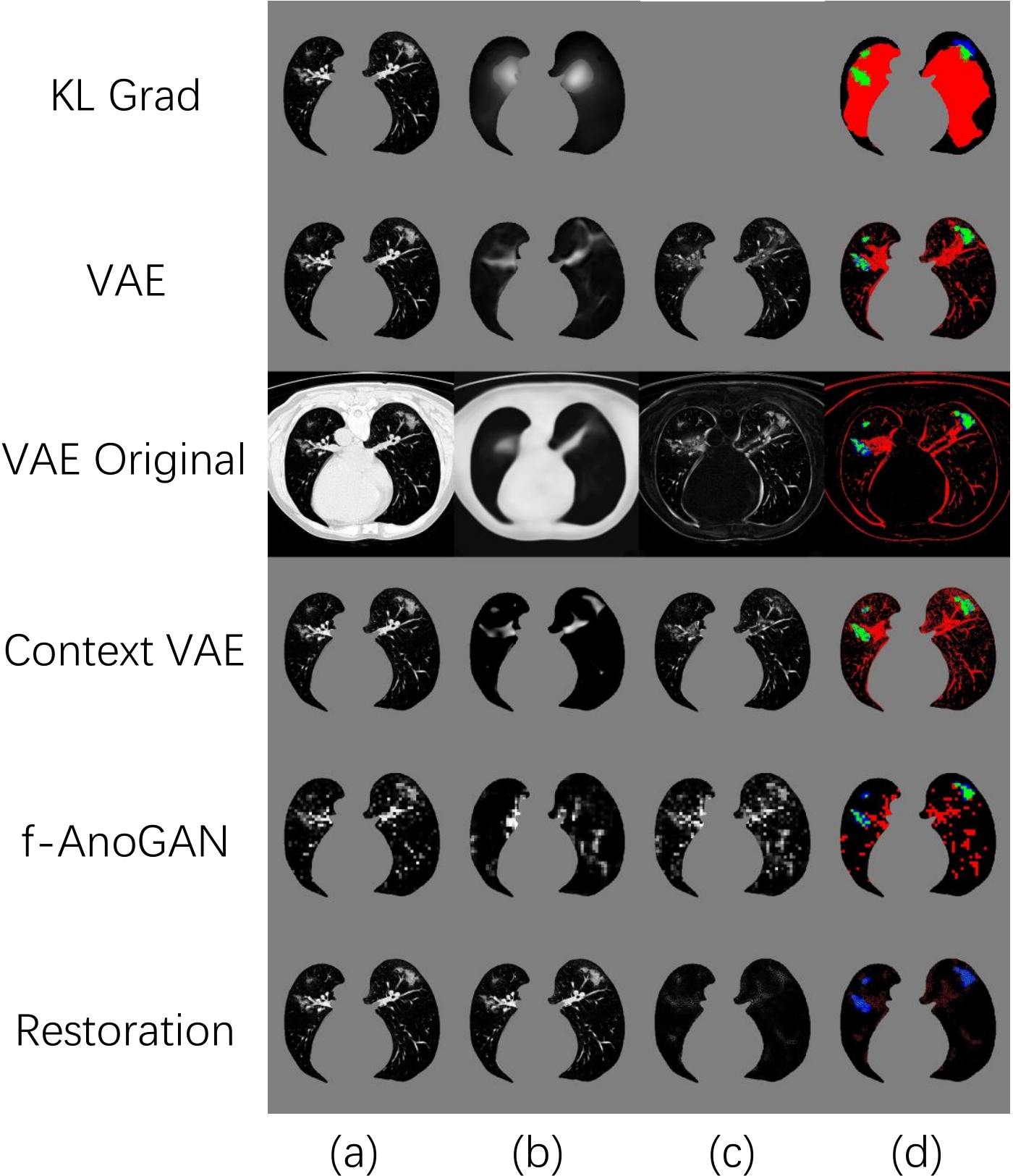}
\caption{Visual results of various UAD methods. (a), and (d) refer to input (after pre-processing), and final results, respectively. The image (b) in the "KL Grad" method means the gradient map of KL loss, while it in the other methods means reconstruction or restoration results. The image (c) of the methods (except for 'KL Grad') means difference map.}
\vspace{-3mm}
\label{fig:baselines}
\end{figure}

On the other hand, the other unsupervised anomaly detection methods have limited power to segment COVID-19 lesion. As shown in Fig. \ref{fig:baselines}, due to the inaccurate reconstructions, the reconstruction-based methods such as VAE \cite{VAE} and f-AnoGAN \cite{f-anogan} can not reconstruct the healthy tissues precisely. On the other hand, the encoder can not make sure to treat the COVID-19 lesion as anomaly, and suppress the lesion in the reconstruction results. These two serious shortcomings result in low COVID-19 segmentation performances, reported in Table \ref{dice}. Compared to other UAD methods, NormNet captures the healthy signals and segments anomalies more precisely.

\subsection{Ablation study}

\subsubsection{Voting} 

To explore the effects of randomness in the training process, we evaluate the performances of the 5 models and thier voting results with  different number of iterations. As shown in Table \ref{iteration}, the performances of the 5 models oscillate as the iteration increases, while the NormNet alleviates this problem through the voting mechanism of 5 models.

\begin{table}[htp]
\centering
 \caption{The Dice scores of five models and voting performance with different number of iterations on Coronacases.} 
 \centering
 \scalebox{0.9}{
 \begin{tabular}{crrrrrr} 

  \toprule 
Iterations & model$_1$ & model$_2$ & model$_3$ & model$_4$ & model$_5$ & voting    \\ 
  \midrule 
1500  &  68.5 & 69.4 & 63.9 & 70.5 & 69.8 & 68.9   \\
2000  &  70.2 & 70.0 & 69.8 & 68.8 & 66.1 & 69.8   \\
2500  &  68.0 & 63.4 & 71.4 & 66.7 & 69.9 & 69.2   \\
  \bottomrule 

 \end{tabular} 
 }
\label{iteration}
\end{table}

\subsubsection{Modules of synthetic `lesion' generator}


The steps of synthetic `lesion' generator can be roughly divided into three parts: Generate shapes ($G_j$ in Section \ref{sec:shapes}), probability maps ($a_i$ in \eqref{Eq:probablity}), and salt noises ($B_i$ in \eqref{Eq:Salt}). To investigate the influence of each part, we train a new NormNet without the corresponding diversity:
\begin{itemize}
    \item \textbf{Fixed shapes ($G_i$):} Generate 5 ellipsoids with radius = 12 for any lung area $H_i$ without any deformation.
    \item \textbf{Fixed probability maps ($a_i$):} Set $a_i = 0.2$.
    \item \textbf{Fixed salt noises ($B_i$):} Set $\sigma_b = 2$ and $\mu_0 = 150$ for synthetic salt noises with the same texture.
\end{itemize}

 As shown in Table \ref{Table:ablation}, the loss of diversities affects the accuracy of the decision boundary and the segmentation performance. Especially, the biggest performance drop in `Fixed $B_i$' prompts that the diverse salt noises make the largest contributions to encourage NormNet to learn tight decision boundary around the normal tissues. 
 
\subsubsection{Hyparameter analysis}

\begin{figure}[htp]
\centering
\centerline{\includegraphics[width=\columnwidth]{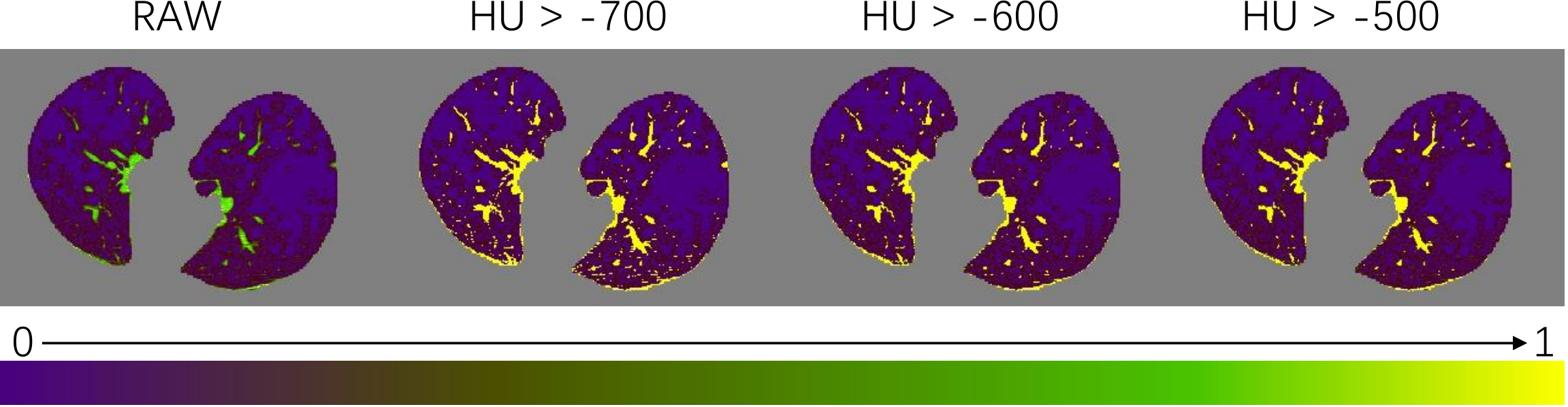}}
\caption{The visualization of masks under different HU thresholds. Many noisy voxels with complex contexts occur when setting the threshold as $T= -700$. We use a colormap for better visualization of the nuances.}
\vspace{-3mm}
\label{fig:HU_threshold}
\end{figure}

\textbf{The threshold of HU:} $T$ is important in our method, since it filters the background noises while trying to keep the pattern complexity at a level that can be effectively managed by the network. On the one hand, if the threshold is too high, our NormNet only segments healthy voxels in a small-scale set, which causes more abnormal voxels missing. On the other hand, if the threshold is too small, some noisy voxels with complex contexts (as shown in Fig. \ref{fig:HU_threshold}) weaken the ability of NormNet to correctly model the normal voxels. As shown in Table \ref{Table:ablation}, the performance drops rapidly when the HU threshold $T = -700$.


\textbf{Hyper-parameters of `lesion' generator:} For the sensitivity analyses, we change the choice of parameters separately\footnote{We mark the experiments with Roman numbers, e.g., (\romannumeral1).}:

\begin{itemize}
    \item \textit{Shape of ellipsoids}: Turn off the elastic-deformation\footnote{Due to the limitation of computation resources, we turn off elastic-deformation in all of the following experiments.} (\romannumeral1) and rotation (\romannumeral2), respectively. 
    \item \textit{Number of ellipsoids}: (\romannumeral3) Generate fewer ellipsoids by changing $N_m,N_s \sim F[5\lambda,10\lambda]$ to $N_m,N_s \sim F[3\lambda,8\lambda]$ and $P_L$ from $0.2\lambda$ to  $0.1\lambda$. (\romannumeral4) Generate more ellipsoids by setting $N_m,N_s$ to $\sim F[7\lambda,12\lambda]$ and $P_L = 0.3\lambda$. 
    \item \textit{Size of ellipsoids}: (\romannumeral5) Select the principal semi-axes of large-size and small-size ellipsoids from $U[3,10]$ and $U[32,64]$ to $U[6,10]$ and $U[32,48]$ , respectively.
    \item \textit{Filtering $a(x)$}: Select the standard deviation $\sigma_a$ from  (\romannumeral6) $U[2,18]$ and  (\romannumeral7) $U[4,12]$, respectively. 
    \item \textit{Scaling $a(x)$}: Set the range of $a_U, a_L$ from $U[0, 0.3]$ to (\romannumeral8) $U[0.5, 0.25]$  and (\romannumeral9) $U[0, 0.35]$, respectively.
    \item \textit{Filtering $b(x)$}: (\romannumeral10) Change the probability values of 0.7 and 0.3 in Eq.~(\ref{Eq:sigma_b}) to 0.5 and 0.5, respectively; (\romannumeral11) Change $U[0.8, 2]$ in Eq.~(\ref{Eq:sigma_b}) to $U[0.6, 2]$; and (\romannumeral12) Change $U[2, 5]$ in Eq.~(\ref{Eq:sigma_b}) to $U[2, 4]$. 
    \item \textit{Scaling $b(x)$}: Set the range of $\mu_0$ from $U[0.4, 0.8]$ to  (\romannumeral13) $U[0.45, 0.75]$ and (\romannumeral9) $U[0.35, 0.85]$, respectively.
\end{itemize}

As shown in Table~\ref{Table:ablation}, the performances of most experiments are stable and greatly outperform other UAD methods. This confirms that there is a wide of range of parameter choices for the `lesion' generator as long as it can produce diverse `lesions' with a balanced probability, forming a rich `lesion' database. Therefore, the decision boundary of the learned NormNet can separate out the distribution of normal tissue, thereby segmenting COVID-19 lesions from normal tissues. 

\textbf{Hyper-parameters of post-processing:} Here, we set up four experiments by individually changing the kernel sizes $k$ and threshold $t$ for both Fig.~\ref{post}(d) and Fig.~\ref{post}(f): $k_d = 7$, $k_f = 9$, $t_d = 0.15$, $t_f = 0.2$. When the hyperparameter makes a small fluctuation, all of the performances are stable.

\subsubsection{Other training strategies}
\footnote{More analyses and visualizations in both training and inference stages can be found in the supplementary material.}
 \textbf{Without removing erroneous edges (`edge' in Table~\ref{Table:ablation}):} Use $M_i = M'_i$ as lung mask in Section~\ref{sec:edge}. Despite our lung-segmenting nnU-Net achieves a high performance in lung segmentation, there are still some false positives around the edge of lung, which appear random and noisy. These noisy textures without consistent patterns confuse the NormNet to capture regular normal textures, which cause the performance drop in Table~\ref{Table:ablation}.
 
\textbf{Directly segmenting healthy areas instead of healthy tissues (`areas' in Table~\ref{Table:ablation}):} There are three types of voxels in the lung area: 1) Plenty of `air' voxels~\cite{ct-lung}, whose intensities are around 0 after clipping with a Hounsfield unit (HU) range of $[-800,100]$ and scaled to $[0,1]$; 2) Healthy tissues; and 3) COVID-19 or synthetic lesions. Here, we redefine the ground-truth $GT_i$ in Eq.~(\ref{Eq:gt}) as $M_i \odot (1 - G)$, which represents `healthy areas' instead of the original `healthy tissues'. In this setting, the NormNet is trained to segment too many low-intensity voxels (voxels of ‘air’) as healthy voxels, rather than focus on healthy tissues whose voxels lie in the high-intensity range. This imbalance limits the power of precisely recognizing those healthy tissues in high intensities range from various anomalies (lesions). Thus, false-positives occur when segmenting COVID-19 CT volumes.

\textbf{Directly segmenting synthetic `lesions' (`lesions' in Table~\ref{Table:ablation}):} Here, we set $GT_i = M_i \odot G$ in Eq.~(\ref{Eq:gt}) to force the NormNet to segment synthetic `lesions' from `air' voxels and healthy tissues directly. However, because there are still differences between synthetic and COVID-19 lesions, the segmentation network has more risk to over-fit the synthetic `lesions'. 
On the contrary, to recognize healthy tissues from plenty of `air' voxels and lesions, the segmentation network must be highly sensitive to healthy tissues. The learned tight decision boundary arising from such sensitivity can be used to segment plenty of anomalies including synthetic and COVID-19 lesions with better generalization, which is the motivation to design the NormNet. The experimental results are shown in Table~\ref{Table:ablation}, in which performance drops in Dice coefficient are clearly observed for the model of learning to directly segment synthetic `lesions'.
 
\section{Conclusions and Discussions}

In this paper, we proposed the NormNet, a voxel-level anomaly modeling network to recognize normal voxels from possible anomalies. A decision boundary for normal contexts of the NormNet was learned by separating healthy tissues from the diverse synthetic `lesions', which can be further used to segment COVID-19 lesions, without training on any labeled data. The experiments on three different COVID-19 datasets validated the effectiveness of the NormNet.

\begin{figure}[htbp]
\centerline{\includegraphics[width=0.8\columnwidth]{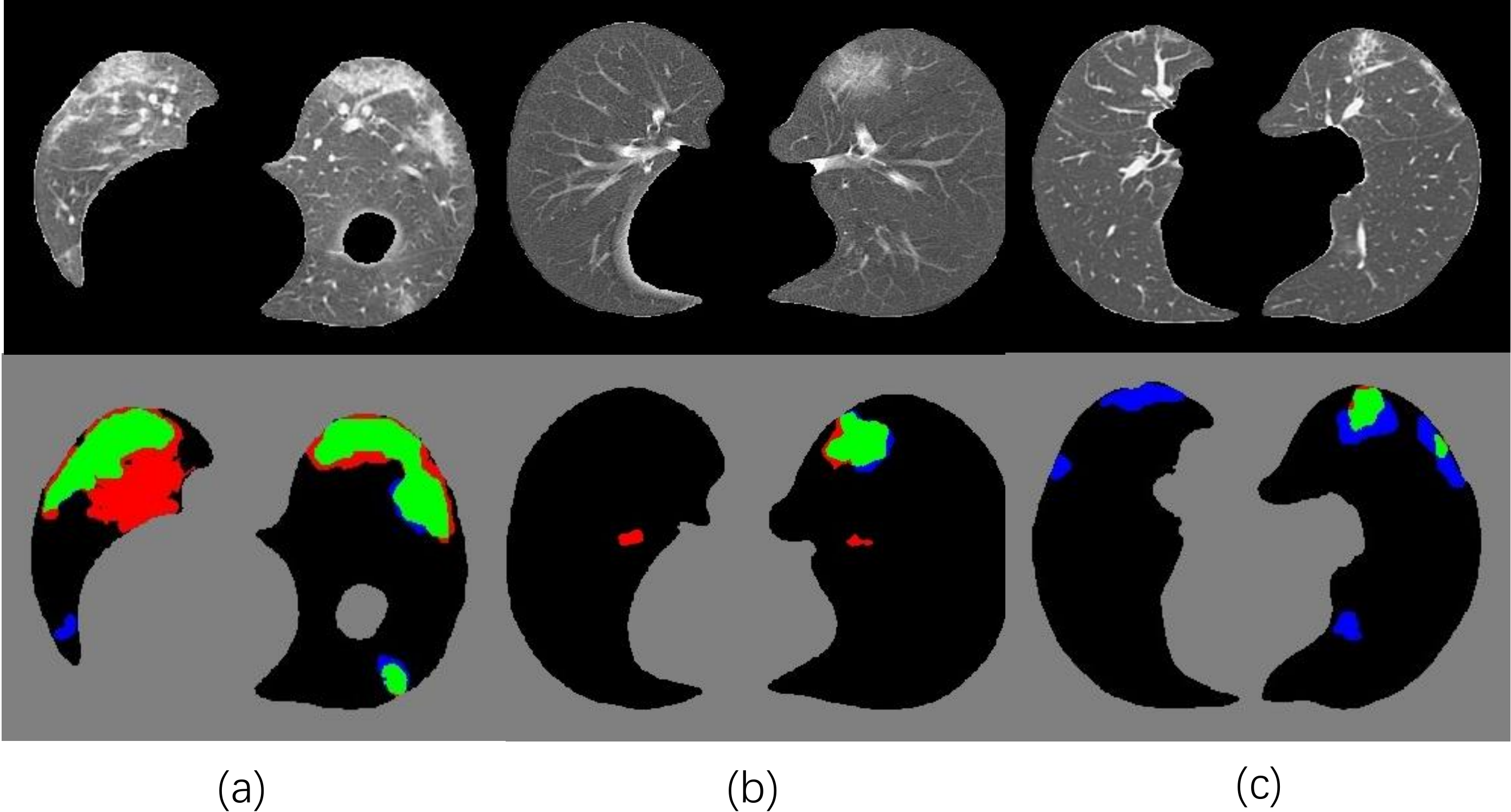}}
\caption{Samples of failure predictions to show the limitation of our method. The red area means false positive while the blue area indicates false negative. }
\vspace{-3mm}
\label{limitation}
\end{figure}

Despite the improvement compared to existing unsupervised anomaly detection methods, there was still a gap between our methods and supervised methods such as nnU-Net \cite{nnunet}. After exploring the failure predictions of our methods, we found that they were divided into three categories:

\begin{enumerate}
    \item The NormNet segments all anomalies such as pulmonary fibrosis (the first row shown in Fig. \ref{limitation}), rather than COVID-19 lesions only.
    \item Gaps between datasets: for example, most of the layer thicknesses in Luna16 dataset are around 1mm. However, in Radiopedia dataset slices were padded together, which generated different contexts. The unseen contexts were treated as anomalies by our NormNet, which resulted in the most of false-positives in Radiopedia.
    \item Our NormNet gave up modeling the noisy patterns in low-intensity range. Although most of lesions can be successfully detected, a small part of lesions with their intensity smaller than $\tau$ were still missed (as shown in the right column of Fig. \ref{limitation}). Segmenting these small lesions also serves as a difficult problem for both supervised methods \cite{tmi-noise} and anomaly detection.
\end{enumerate}

For a better performance on COVID-19 segmentation, we plan to extend our method to address the above limitations mainly in the following three aspects: 1) Modeling more `non-COVID-19' contexts including other diseases; and 2) Exploring a better way of modeling low-intensity normal voxels as much as possible by mitigating the impact of noise with an array of denoising methods. 3) Creating a more effective synthetic ‘lesions’
generator for network learning by exploring different generation schemes, such as using a deeper hierarchy and a universal generation \cite{u2net} by investigating cross-anatomy or even cross-modality possibilities. 4) Exploring the idea of metric learning such as Deep SVDD \cite{dsvdd} to get tighter decision boundary.  

Beyond COVID-19 lesion segmentation in Lung CT, we believe that it is possible to extend the NormNet to other modalities (e.g. MRI) by defining a similar proxy task, such as denoising or inpainting, etc. The NormNet can be learned to `recover' the polluted healthy texture back to normal if the contexts of healthy tissues are sufficiently captured. We are going to investigate this direction in future.


\end{document}


\title{Label-Free Segmentation of COVID-19 Lesions in Lung CT \\
\textit{Supplementary Material} }
\maketitle

\vspace{-25mm}

\section{Hyper-parameter analyses and motivations of Lesion generator }

\begin{figure}[h]
\centering
\includegraphics[width=0.6\columnwidth]{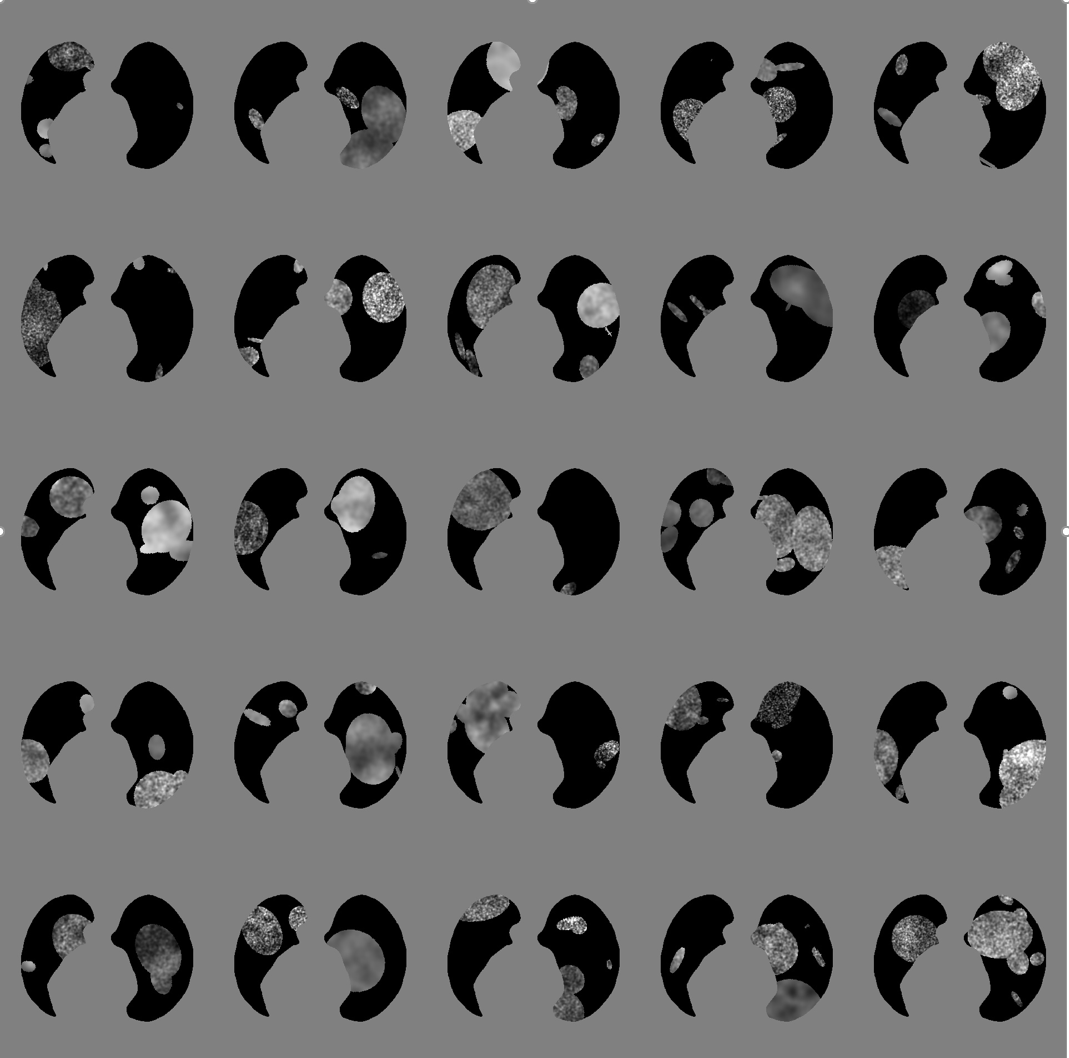}
\caption{Several fake `lesions' synthesized from the generator.}
\label{Fig:revision-fake-lesion}
\end{figure}

As discussed in the manuscript, a rich `lesion' database encourages the NormNet to recognize healthy tissues from diverse synthetic `lesions', resulting in the desired tight decision boundary. Hence, the motivation to `lesions' generator is to \textbf{make the synthetic `lesions' as diverse as possible}, and we do not aim at \textit{fitting} these parameters to the COVID-19 lesions from a specific dataset. It should be noted that our operators and the choices of parameters are far from prefect; nevertheless, we find it is empirically effective. Fig.~\ref{Fig:revision-fake-lesion} presents several examples of synthesized lesions.

In this subsection, we provide the visualization and segmentation performance to show the effect of each parameter. 
Based on the notion of \textit{generating diverse `lesions' evenly}, we experimentally choose variants of hyperparameters that are roughly divided into the following categories: the shape of ellipsoids, the number of ellipsoids, the size of ellipsoids, the filtering for probability function $a(x)$, the scaling for probability function $a(x)$, the filtering for salt noise $b(x)$, and the scaling for salt noise $b(x)$:


\begin{figure}[h]
\centering
\includegraphics[width=0.9\columnwidth]{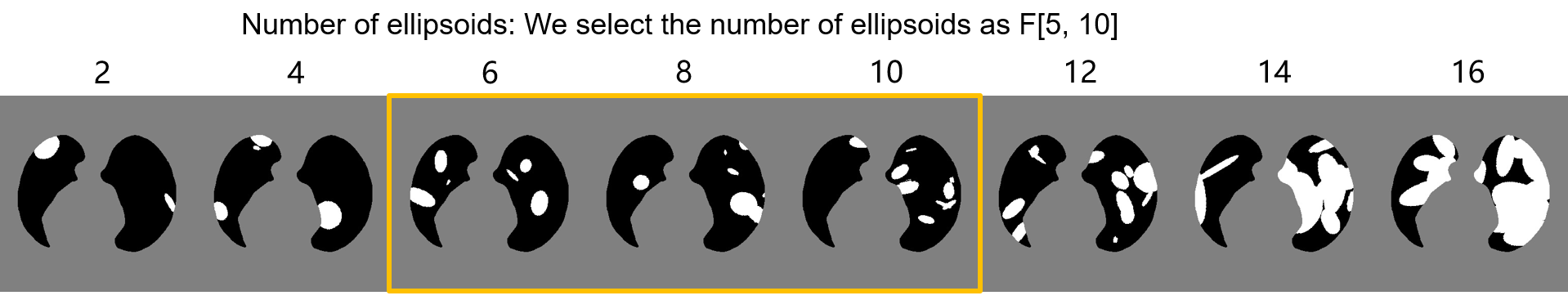}
\caption{The effect by varying the number of ellipsoids. We aim at generating diverse anomalies evenly and keeping the number of ellipsoids at a proper level.}
\label{Fig:revision-num-ellipsoid}
\end{figure}
    
    
\begin{figure}[h]
\centering
\includegraphics[width=0.6\columnwidth]{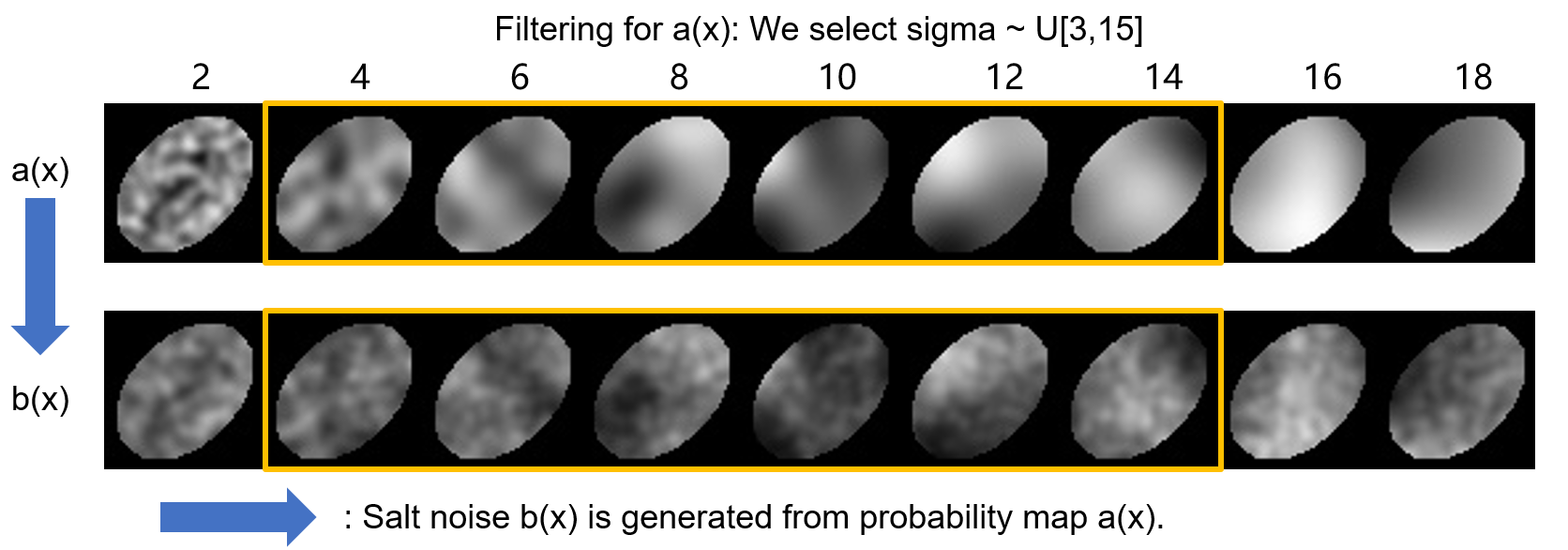} (a)
\includegraphics[width=0.3\columnwidth]{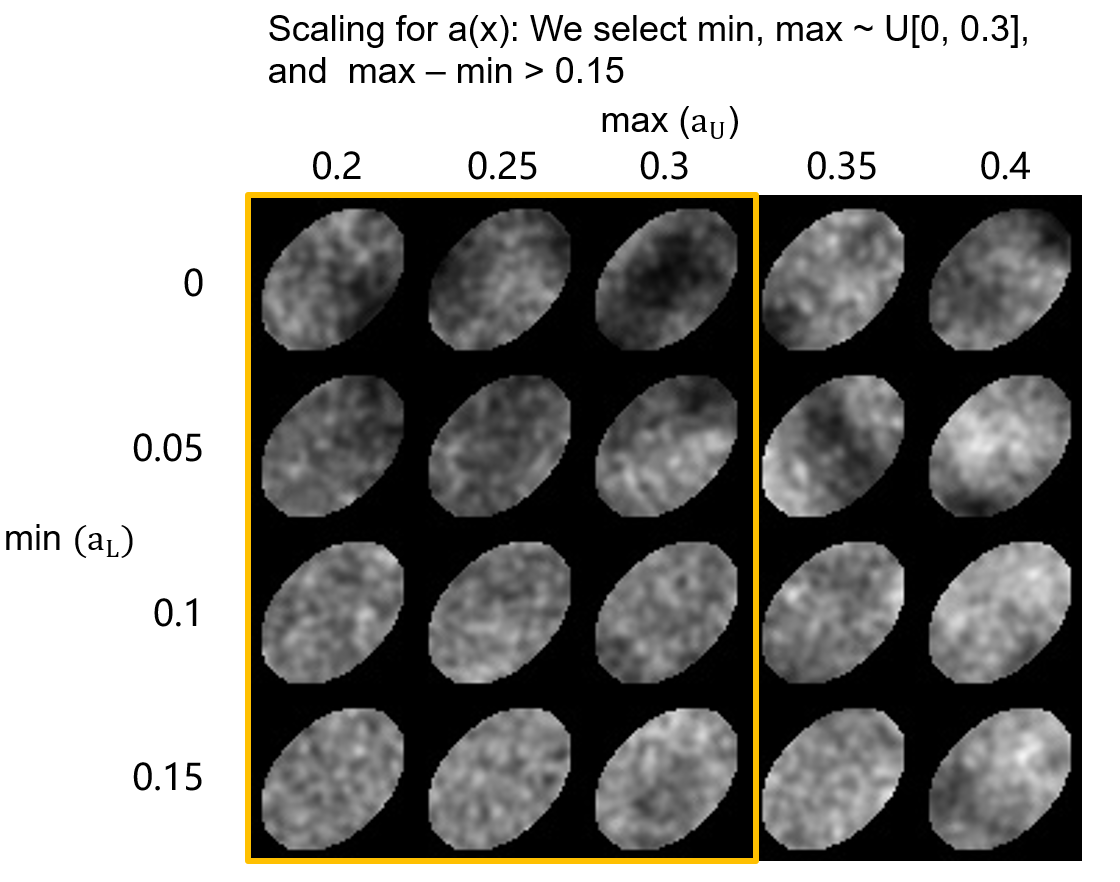} (b)
\caption{(a) The effect by varying $\sigma_a$ that is used for filtering $a(x)$. (b) The effect by varying $a_U$ and $a_L$ that are used for scaling $a(x)$.}
\label{Fig:revision-filter1}
\end{figure}

    
    


\begin{figure}[H]
\centering
\includegraphics[width=0.6\columnwidth]{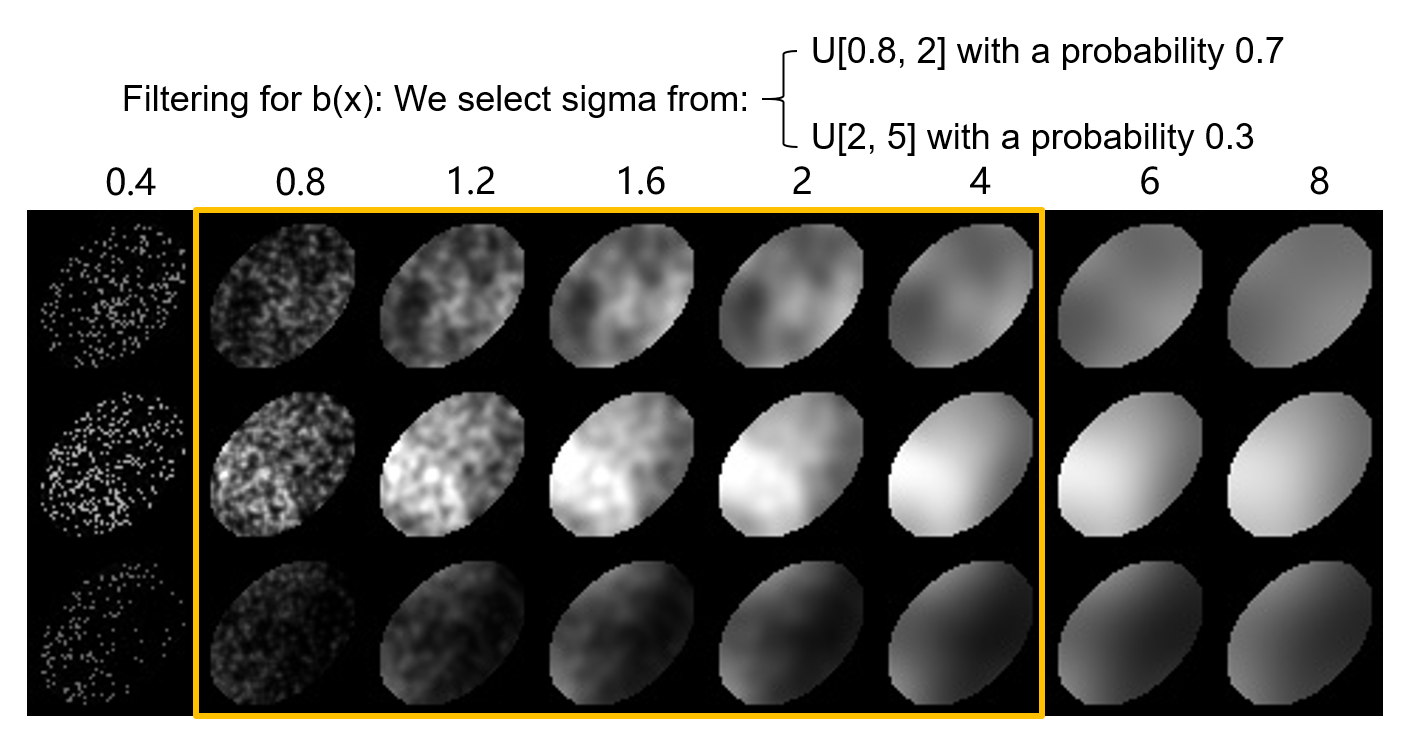}
\caption{The effect by varying  $\sigma_b$ that is used for filtering $b(x)$.}
\label{Fig:revision-temp}
\end{figure}

    
\begin{figure}[H]
\centering
\includegraphics[width=0.8\columnwidth]{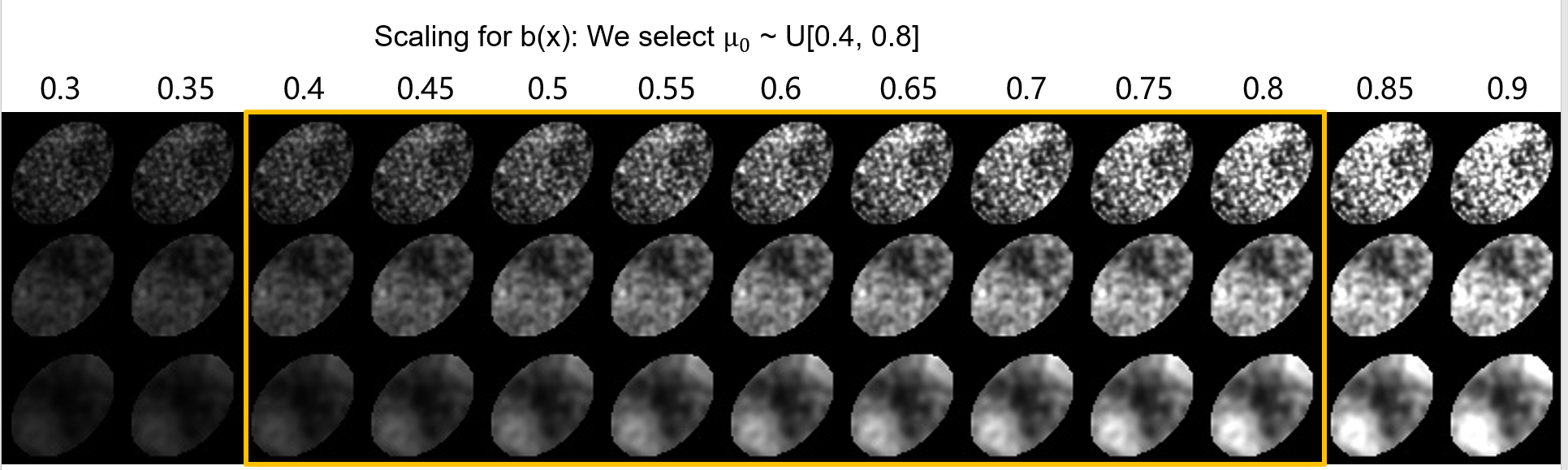}
\caption{The effect of $\mu_0$ when scaling for b(x). We aim at generating diverse anomalies evenly.}
\label{Fig:revision-filter3}
\end{figure}

\section{Analyses and visualizations of other training strategies}

\subsection{Without removing erroneous edges}

Our lung-segmenting nnU-Net achieves a high performance in lung segmentation (dice $>$ .98). However, there are still some false positives around the edge of lung. They are caused by the false predictions of nnU-Net, which appear random and noisy. These noisy textures without consistent patterns confuse the NormNet to capture regular normal textures. In order to make NormNet pay more attention to the normal tissues in thorax, we remove these false-positive edges based on prior experience. 

\begin{figure}[H]
\centering
\centerline{\includegraphics[width=0.5\columnwidth]{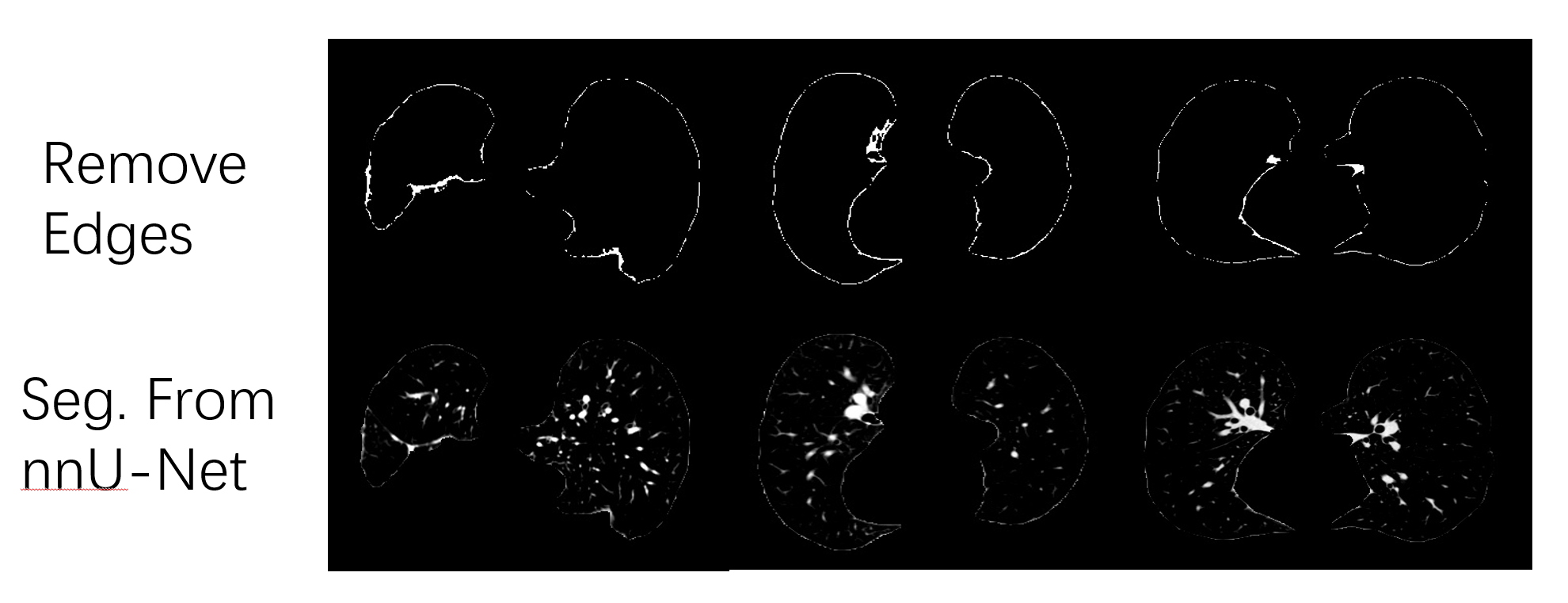}
}
\caption{Visualization of lung segmentation from nnU-Net and the edges we removed.}
\label{Fig:Revision-R2C6}
\end{figure}


\subsection{Directly segment healthy regions instead of healthy tissues}

\begin{figure}[H]
\centering
\centerline{\includegraphics[width=0.6\columnwidth]{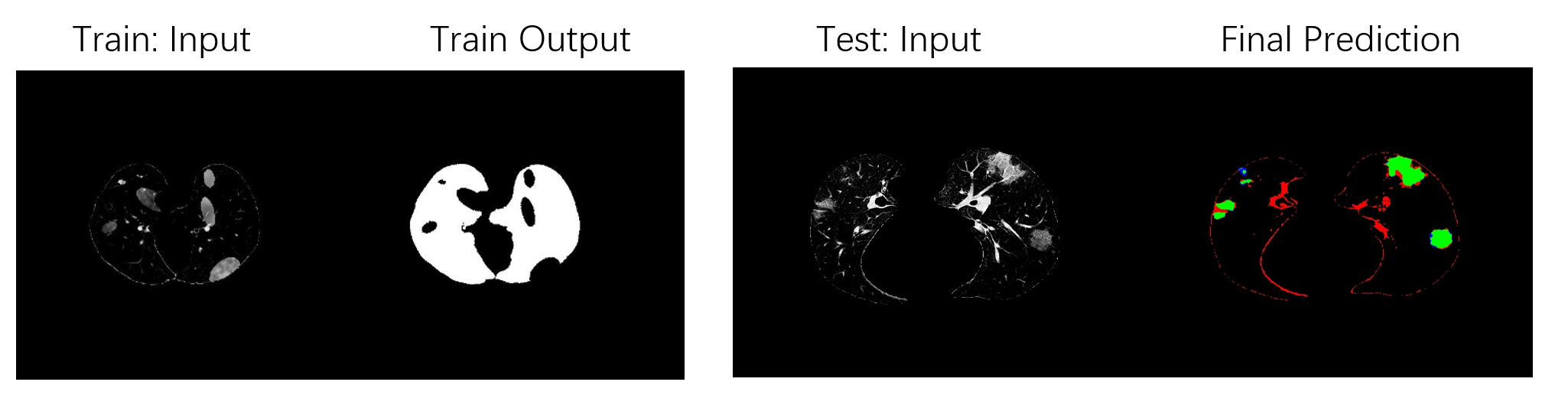}
}
\caption{Visualization of the training and testing pairs from the experiments ``segment healthy regions''.}
\label{Fig:Revision-R2C3}
\end{figure}

In this experiment, we set $GT_i = M\odot(1-G) $. As shown in Fig.~\ref{Fig:Revision-R2C3}, the NormNet is trained to segment too many low-intensity voxels (voxels of `air' in thorax) as healthy regions, rather than focus on healthy tissues, which limits the power of recognizing those healthy tissues in high intensities from various anomalies (lesions). Thus, false-positives occur when segmenting COVID-19 CT volumes. 

\subsection{Using Fig. 4(g) as prediction}

Some low-intensity noisy patterns appears in Fig. 4(f)\footnote{Fig. 4 is the Fourth figure (illustration of post-processing) in the manuscript.}. Directly treat those noisy patterns as anomalies will decrease the precision score. So we choose to find back the lesion information (in the low-intensity range) guided by the accurate lesions located in the high-intensity range. The ablation study can be found in Table~\ref{Table:revision-g}.

\begin{table}[htp]
\centering
 \caption{The performance of using Fig. 4(g) as prediction. DSC: Dice coefficient, PSC: precision, SEN: sensitivity} 
 \centering
 \small
\scalebox{1}{
 \begin{tabular}{l|rrr|rrr} 
  \toprule 
\multirow{2}{*}{Datasets} & \multicolumn{3}{c|}{Using Fig. 4(g) as prediction} & \multicolumn{3}{c}{Original NormNet} \\ 
 &  DSC & PSC & SEN &  DSC & PSC & SEN      \\ 
  \midrule 
Coronacase & 67.3$\pm$14.3 & 75.7$\pm$9.67 & 67.9$\pm$21.9 & 69.8$\pm$15.2 &82.1$\pm$8.92 &66.2$\pm$22.2  \\
Radiopedia & 62.5$\pm$13.7 & 59.1$\pm$18.4 & 70.8$\pm$13.1 & 59.3$\pm$16.9 & 58.3$\pm$18.0 &65.6$\pm$18.7\\
UESTC & 59.8$\pm$20.2 & 57.4$\pm$27.0 & 79.8$\pm$17.6 & 61.4$\pm$19.4 &61.3$\pm$26.1 &77.6$\pm$19.6 \\
  \bottomrule 
 \end{tabular} 
}
\label{Table:revision-g}
\end{table}


